\documentclass[preprint,12pt]{elsarticle}

\usepackage{hyperref}
\usepackage{graphicx}
\usepackage{subcaption}
\usepackage{amssymb}
\usepackage{amsmath}
\usepackage{multirow}
\usepackage{booktabs}

\usepackage[utf8]{inputenc}

\usepackage{xcolor}
\usepackage{xargs}
\usepackage[colorinlistoftodos,textsize=footnotesize]{todonotes}

\newcommandx{\unsure}[2][1=]{\todo[linecolor=red,backgroundcolor=red!25,bordercolor=red,#1]{#2}}
\newcommandx{\change}[2][1=]{\todo[linecolor=blue,backgroundcolor=blue!25,bordercolor=blue,#1]{#2}}
\newcommandx{\info}[2][1=]{\todo[linecolor=OliveGreen,backgroundcolor=OliveGreen!25,bordercolor=OliveGreen,#1]{#2}}

\newcommand{\ten}[1]{\ensuremath{\mathbf{#1}}}

\usepackage{lineno}
\usepackage{tikz}
\usetikzlibrary{shapes.geometric, arrows}
\usetikzlibrary{arrows.meta}
\usetikzlibrary{calc,quotes,angles}
\usetikzlibrary{shapes,snakes}
\usetikzlibrary{arrows.meta}
\usetikzlibrary{calc,quotes,angles}
\tikzset{%
  >={Latex[width=2mm,length=2mm]},
            base/.style = {rectangle, rounded corners, draw=black,
                           minimum width=4cm, minimum height=1cm,
                           text centered, font=\sffamily},
  activityStarts/.style = {base, fill=blue!30},
       startstop/.style = {base, fill=red!30},
    activityRuns/.style = {base, fill=green!30},
         process/.style = {base, minimum width=2.5cm, fill=orange!15,
                           font=\ttfamily},
}
\journal{}

\begin{document}

\begin{frontmatter}

  \title{An improved non-reflecting outlet boundary condition for
    weakly-compressible SPH}
  \author[IITB]{Pawan Negi\corref{cor1}}
  \ead{pawan.n@aero.iitb.ac.in }
  \author[IITB]{Prabhu Ramachandran}
  \ead{prabhu@aero.iitb.ac.in}
  \author[IITB]{Asmelash Haftu}
  \ead{asmelash.a@aero.iitb.ac.in}
\address[IITB]{Department of Aerospace Engineering, Indian Institute of
  Technology Bombay, Powai, Mumbai 400076}

\cortext[cor1]{Corresponding author}

\begin{abstract}
  Implementation of an outlet boundary condition is challenging in the context
  of the weakly-compressible Smoothed Particle Hydrodynamics method. We
  perform a systematic numerical study of several of the available techniques
  for the outlet boundary condition. We propose a new hybrid approach that
  combines a characteristics-based method with a simpler frozen-particle
  (do-nothing) technique to accurately satisfy the outlet boundary condition
  in the context of wind-tunnel-like simulations. In addition, we suggest some
  improvements to the do-nothing approach. We introduce a new suite of test
  problems that make it possible to compare these techniques carefully. We
  then simulate the flow past a backward-facing step and circular cylinder.
  The proposed method allows us to obtain accurate results with an order of
  magnitude less particles than those presented in recent research. We provide
  a completely open source implementation and a reproducible manuscript.
  \end{abstract}


\end{frontmatter}


\section{Introduction}
\label{sec:intro}

The Smoothed Particle Hydrodynamics (SPH) method was independently introduced
by \citet{monaghan-gingold-stars-mnras-77}, and \citet{lucy77} for simulation
of astrophysical problems. Ever since, many SPH schemes have been introduced
to solve a variety of fluid flow and elastic-dynamics problems (see
\cite{monaghan-review:2005} for a review). \citet{sph:fsf:monaghan-jcp94}
introduced the weakly-compressible SPH (WCSPH) to deal with incompressible
fluids like water. An equation of state is introduced to relate the pressure
to the density. There are two common problems with the WCSPH schemes. The
first is the presence of particle disorder which reduces the accuracy of the
scheme and the second is the presence of large pressure oscillations due to
the stiff equation of state. Particle disorder can be ameliorated by the use
of the Transport Velocity Formulation (TVF) \cite{Adami2013,zhang_hu_adams17}
or by using particle
shifting~\cite{diff_smoothing_sph:lind:jcp:2009,fickian_smoothing_sph:skillen:cmame:2013}.
The pressure oscillations can be reduced by using a density
smoothing~\cite{wcsph-state-of-the-art-2010}, or by use of the $\delta$-SPH
formulation~\cite{antuono-deltasph:cpc:2010}. In the present work, we have
used the Entropically Damped Artificial Compressibility SPH (EDAC SPH)
\cite{edac-sph:cf:2019} method that introduces a pressure evolution equation
that damps any pressure oscillations. In addition to the WCSPH schemes
discussed above, there are also a family of truly Incompressible SPH
schemes~\cite{sph:psph:cummins-rudman:jcp:1999,isph:shao:lo:awr:2003,isph:hu-adams:jcp:2007}
(ISPH). These schemes solve for a pressure-Poisson equation to find a suitable
pressure distribution. These schemes require that a large, sparse system of
linear equations be solved in order to compute the pressure.

Despite the many developments in the SPH method, there are some challenges in
implementing accurate non-reflecting boundary conditions (NRBC) with the
weakly-compressible formulations. One significant objective in implementing
inlet and outlet boundary conditions is to let the pressure and velocity
fluctuations pass out of the domain without affecting the internal particles.
\citet{Lastiwka2009:nonrefbc} addressed this by extrapolating properties from
within the fluid. To obtain first order consistency near the inlet and outlet,
the reproducing kernel particle method given by \citet{Liu1995:rkpm} is used.
Within the fluid, the corrected gradient given by \citet{bonet_lok:cmame:1999}
is used for accurate results. Any perturbations are passed out of the domain
using characteristic variables and carefully chosen boundary conditions based
on these characteristic variables.

\citet{FEDERICO201235} proposed freezing the properties of the fluid particles
in the outlet. The outlet particles are advected with the frozen velocity.
\citet{MARRONE2013456} utilized the approach suggested by
\citet{FEDERICO201235} and \citet{Lastiwka2009:nonrefbc} to simulate flows
around bluff-bodies for a wide range of Reynolds numbers.
\citet{MOLTENI201378} proposes using a sponge layer in order to absorb waves
coming from the domain in order to implement a non-reflective boundary and
tested it on water waves inside a tank. In a method suggested by
\citet{khorasanizade-open-boundary-sph:ijnmf-2015}, the fluid is divided into
multiple sections perpendicular to the flow and the values from these zones
are used to impose natural boundary conditions (zero-gradients of properties)
to the inlet/outlet.

Recently, \citet{ALVARADORODRIGUEZ2017177} modified the NRBCs proposed by
\cite{JIN1993239} for SPH. \citet{open_bc:tafuni:cmame:2018} proposes the use
of ghost (or buffer) particles for inlet/outlet particles and use the higher
order interpolation scheme of \citet{LIU200619} to extrapolate the property
using a Taylor series expansion. Their approach allows them to treat the
outlet and inlet buffer particles in the same way. \citet{pwang_2019} use the
characteristic wave propagation velocity and perform Lagrange interpolation in the time
domain to correctly implement NRBCs to simulate an under water blast in a
small domain.

In the context of ISPH schemes, \citet{HOSSEINI20117473} suggested a
rotational pressure correction scheme in order to extrapolate pressure to the
inlet or outlet and thereby impose natural boundary conditions. At the outlet,
the last layer of fluid is copied up to a sufficient distance to ensure kernel
support for the fluid particles. \citet{PAHAR2017464} satisfy a
divergence-free condition for the inlet and outlet by solving a
pressure-Poisson equation along with the fluid particles.
\citet{MONTELEONE20179} investigated a novel approach in which only pressure
boundary conditions were prescribed and velocity profiles are allowed to
change according to it.

In the present work we focus on weakly-compressible schemes. It is clear that
the method proposed by \citet{Lastiwka2009:nonrefbc} is ideal when one wishes
to extrapolate properties from the fluid. This is most useful for inlets where
one needs to extrapolate the pressure from the fluid into the inlet and
prescribe the inlet velocity alone. However, for outlets, it is not clear
which one of these methods is ideal for bluff body simulations. We find that
there are a few important considerations that are not fully discussed in any
of the earlier studies. Specifically, many realistic flows involving an outlet
will have large vortices leaving the domain. These vortices involve both a
pressure and velocity gradient. It is important that any outlet boundary
condition not destroy these structures as doing so would affect the vortices
upstream. These are typically handled by simply increasing the domain but this
may not be needed if the outlet is carefully implemented. WCSPH schemes
constantly generate pressure waves. These may be severe if the bodies are
oscillating and this would introduce additional pressure waves which should be
propagated out of the domain without vitiating any physical gradients like
those due to vortices.

Efficiently testing an SPH outlet implementation in the context of the above
issues is critical. Doing so using a flow past cylinder benchmark is
inefficient. We propose a suite of simple and efficient test problems that
allow us to systematically investigate the boundary conditions. The benchmarks
are the simple one-dimensional benchmark proposed by
\citet{Lastiwka2009:nonrefbc}, a two-dimensional wave, a free-vortex advecting
with a mean flow, and a ramp inlet.

We implement the following boundary conditions and test them with the above
benchmark problems and bring out the relative merits of each. The methods we
implement are,
\begin{itemize}
\item a simple do-nothing boundary
  condition~\cite{khorasanizade-open-boundary-sph:ijnmf-2015,ALVARADORODRIGUEZ2017177}
  where the particle properties are frozen. We propose an improvement to this
  method .
\item extrapolating the fluid properties to the outlet as proposed by
  \citet{open_bc:tafuni:cmame:2018} using a higher order interpolation.
\item propagate the properties of the fluid into the outlet using the
  method of characteristic (MOC)~\citet{Lastiwka2009:nonrefbc}.
\item a new hybrid approach that combines the do-nothing and MOC methods.
\end{itemize}

Based on our careful study, we see that all the existing methods have some
difficulties. The hybrid method uses the best features of the available
methods and performs much better with our test problems. The proposed
modification to the traditional do-nothing also produces fairly good results
and is very easy to implement. We finally apply these to the flow past a
backward-facing step and a cylinder for Reynolds numbers in the range 20-200.
We present the results in the entire computational domain showing the
effectiveness of our implementation. The new boundary conditions allow us to
obtain reasonable results with an order of magnitude fewer particles than
previous results.

We use the open source PySPH~\cite{PR:pysph:scipy16,pysph} framework for our
simulations. Furthermore, in the interest of reproducible research, every
figure presented in the results section of this manuscript is
automated~\cite{pr:automan:2018} and the source code is made available at
\url{https://gitlab.com/pypr/inlet_outlet}. In the next section, we describe
the SPH scheme we employ in some detail. Section \ref{sec:bc}, discusses the
different techniques used to implement the outlet boundary conditions.
Section~\ref{sec:results} introduces the new test problems and compares the
different boundary condition implementations.

\section{The SPH method}
\label{sec:sph}

In the present work, the EDAC (Entropically Damped Artificially Compressible)
SPH scheme~\cite{edac-sph:cf:2019} is used to simulate incompressible fluid
flow. The EDAC scheme uses a pressure evolution equation that is similar to
the continuity equation but also contains a pressure damping term which reduces
oscillations. The basic equations are the momentum equation,
\begin{align}
	\label{eq:mom-newt}
	\frac{d \ten{u}}{d t} &= -\frac{1}{\rho} \nabla p +
	\nu \nabla^2 \ten{u},
\end{align}
where $\ten{u}$ is the velocity of the fluid, $p$ is the pressure, and $\nu$
is the kinematic viscosity of the fluid. The EDAC pressure equation is given
as,
\begin{equation}
 	\label{eq:p-evolve}
	\frac {d p}{d t} = - \rho c _ {s}^{2} \operatorname {div} (
	\mathbf {u}) + \nu_{edac} \nabla ^{2} p,
\end{equation}
where $c_s$ is the speed of sound, and the second term in the right hand side
is the damping term and the viscosity used there is chosen as,
\begin{equation}
  \label{eq:edac-nu}
  \nu_{edac} = \frac{\alpha h c_s}{8}.
\end{equation}
$\alpha$ is chosen as 0.5, $h$ is the SPH kernel smoothing length which is
discussed further below and $c_s$ is chosen such that $c_s = 10\ u_{\max}$
where $u_{\max}$ is an estimated maximum speed in the flow.

The EDAC SPH formulation~\cite{edac-sph:cf:2019} comes in two flavors. As we
are primarily solving problems without a free surface in this work, we use the
EDAC TVF formulation which employs the Transport Velocity Formulation of
\cite{Adami2013} along with the EDAC equation for evolving pressure,
equation~\eqref{eq:p-evolve}. This formulation ensures that the particle
distribution is uniform through the use of a background pressure.

Particle volume for a particle $i$ is evaluated using $m_i/\rho_i$ where
$\rho_i$ is evaluated using the summation density,
\begin{equation}
	\label{eq:summation-density}
	\rho_i = \sum_j m_j W_{ij},
\end{equation}
where $W_{ij} = W(|\ten{r_i} - \ten{r_j}|, h)$ is the kernel function chosen
for the SPH discretization and $h$ is the kernel radius parameter. The
summation is over all the neighbors of particle $i$. In this paper, the
quintic spline kernel is used, which is given by,
\begin{equation}
	\label{eq:quintic-spline}
	W(q) = \left \{
	\begin{array}{ll}
		\alpha_2 \left[ {(3-q)}^5 - 6{(2-q)}^5 + 15{(1-q)}^5 \right],\
		& \textrm{for} \ 0\leq q \leq 1,\\
		\alpha_2 \left[ {(3-q)}^5 - 6{(2-q)}^5 \right],
		& \textrm{for} \ 1 < q \leq 2,\\
		\alpha_2 \ {(3-q)}^5 , & \textrm{for} \ 2 < q \leq 3,\\
		0, & \textrm{for} \ q>3,\\
	\end{array} \right.
\end{equation}
where $\alpha_2 = 7/(478\pi h^2)$ in two-dimensions, and $q=|\ten{r}|/h$.

The present work utilizes a number density based formulation as discussed in
\cite{edac-sph:cf:2019}. The resulting discretized momentum equation is as
follows:
\begin{equation}
	\label{eq:tvf-momentum}
	\begin{split}
		\frac{\tilde{d} \ten{u}_i}{d t} = \frac{1}{m_i} \sum_j \left( V_i^2 +
		V_j^2 \right) & \left[ - \tilde{p}_{ij} \nabla W_{ij} +
		\frac{1}{2}(\ten{A}_i + \ten{A}_j) \cdot \nabla W_{ij} \right . \\ &
		\left .  + \tilde{\eta}_{ij} \frac{\ten{u}_{ij}}{(r_{ij}^2 + \eta
			h_{ij}^2)} \nabla W_{ij}\cdot \ten{r}_{ij} \right] + \ten{g}_i,
	\end{split}
\end{equation}
where $\ten{A} = \rho \ten{u}(\ten{\tilde{u}} - \ten{u})$, $\ten{\tilde{u}}$
is the advection or transport velocity and $\frac{\tilde{d}}{dt}$ is the
material derivative associated with this transport velocity. $\ten{r}_{ij} =
\ten{r}_i - \ten{r}_j$, $\ten{u}_{ij} = \ten{u}_i - \ten{u}_j$, $h_{ij} = (h_i
+ h_j)/2$, $\eta=0.01$, $ V_i = \frac{1}{\sum_j W_{ij}} $, and
$\tilde{\eta}_{ij} = \frac{2 \eta_i \eta_j}{\eta_i + \eta_j}$, where $\eta_i =
\rho_i \nu_i$. An average pressure is subtracted to reduce errors in the
pressure gradient. The average pressure is found as,
\begin{equation}
	\label{eq:pavg}
	p_{\text{avg}, i} = \sum_{j=1}^{N_i} \frac{p_j}{N_i},
\end{equation}
where $N_i$ are the number of neighbors for the particle $i$ and includes both
fluid and boundary neighbors. This average pressure is used to define
$\tilde{p}_{ij}$ as,
\begin{equation}
	\label{eq:tvf-p-ij-basa}
	\tilde{p}_{ij} =
	\frac{\rho_j (p_i-p_{avg, i}) + \rho_i (p_j - p_{avg, i})}{\rho_i + \rho_j}.
\end{equation}

The EDAC pressure evolution equation (\ref{eq:p-evolve}) is discretized
using a similar approach to the momentum equation as,
\begin{align}
	\label{eq:p-edac}
	\frac{d p_i }{dt} &=  \sum_j \frac{m_j \rho_i}{\rho_j} c_s^2 \ \ten{u_{ij}}
	\cdot \nabla W_{ij} + \frac{(V_i^2 + V_j^2)}{m_i} \tilde{\eta}_{ij}
	\frac{p_{ij}}{(r_{ij}^2 + \eta h_{ij}^2)} \nabla W_{ij}\cdot \ten{r}_{ij},
\end{align}
where $p_{ij} = p_i - p_j$.

The particles move using the transport velocity as,
\begin{equation}
	\label{eq:advection}
	\frac{d \ten{r}_i}{dt} = \ten{\tilde{u}}_i.
\end{equation}
The transport velocity is obtained from the momentum velocity $\ten{u}$ at
each time step using,
\begin{equation}
	\label{eq:transport-vel}
	\ten{\tilde{u}}_i(t + \delta t) = \ten{u}_i(t) +
	\delta t \left(
	\frac{\tilde{d} \ten{u}_i}{dt}
	- \frac{p_b}{m_i} \sum_j \left( V_i^2 + V_j^2 \right)
	\nabla W_{ij}
	\right),
\end{equation}
where $p_b$ is the background pressure. We choose the timestep and other
parameters as discussed in \cite{edac-sph:cf:2019}. The solid wall boundary
conditions are implemented as discussed in \cite{Adami2012,edac-sph:cf:2019}
and use a layer of ghost particles inside the solid. The pressure and velocity
of the fluid is suitably projected on the solid. While we have employed the
EDAC SPH scheme in our computations, we could have employed any WCSPH-based
scheme for the purposes of this study.

\section{Boundary conditions }
\label{sec:bc}

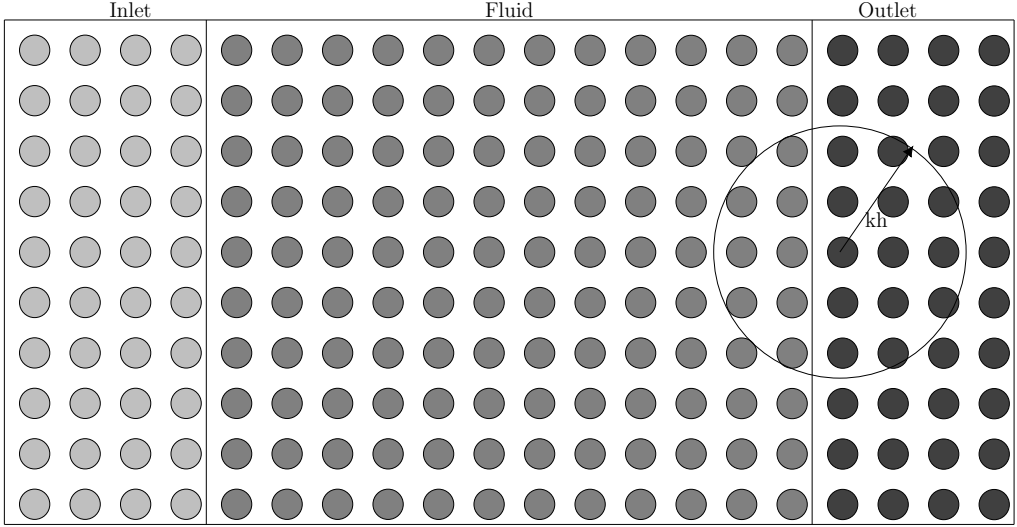
\begin{figure}[h!]
  \centering
  \resizebox{\linewidth}{!}{
    \begin{tikzpicture}[x=1cm, y=1cm, line width=0.1mm]
    \pgfmathsetmacro{\l}{20};
    \pgfmathsetmacro{\w}{10};

    \draw[-] (0, 0) -- (\l, 0);
    \draw[-] (\l, 0) -- (\l, \w);
    \draw[-] (\l, \w) -- (0, \w);
    \draw[-] (0, \w) -- (0, 0);

    \draw[-] (0.2*\l,0) -- (0.2*\l,\w);
    \draw[-] (0.8*\l,0) -- (0.8*\l,\w);

    \foreach \x in {1,...,20}
        \foreach \y in  {1,...,10}
            {
              \pgfmathsetmacro{\res}{ifthenelse(\x<5,"lightgray", ifthenelse(\x<17,"gray","darkgray")}
              \pgfmathsetmacro{\CosValue}{1/cos(\x)}
              \draw[fill=\res] (\x-0.5+\CosValue*0.1,\y-0.5-\CosValue*0.1) circle (0.3);
    };

    \node[yshift=10.2cm, xshift = 2.5cm] {Inlet};
    \node[yshift=10.2cm, xshift = 10cm] {Fluid};
    \node[yshift=10.2cm, xshift = 17.5cm] {Outlet};
    \draw (16.55, 5.40) circle(2.5);
    \draw[->] (16.55, 5.40) -- node[yshift=-0.4cm, xshift=0.0cm]{kh}(18.00, 7.5);
\end{tikzpicture}
  }
  \caption{Sketch of the inlet, fluid, and outlet particle arrangement.  The
    support for one outlet particle is also shown.}
  \label{fig:io_schematic}
\end{figure}

In this work we are interested in simulating incompressible flow and in all of
our test problems we have a prescribed velocity at the inlet. In order for a
fluid particle with support radius $h$ to have full support, one requires
outlet/inlet particles. Fig.~\ref{fig:io_schematic}, shows a schematic for the
particles at the inlet, outlet, and fluid. The particle properties at the
inlet and outlet are evaluated using those of the fluid. As described in the
previous section, the EDAC SPH scheme employs a pressure evolution equation
that is not directly related to the fluid density. We therefore extrapolate
pressure from the fluid to the inlet using the mirroring technique as
described in Section \ref{subsec:mirror}. At the outlet, one needs to
determine values of both the velocity and pressure. In this paper, we first
evaluate the different existing approaches for implementing outlets and
propose improvements in order to simulate NRBCs. In the following subsections,
we describe the methods that we implement.

\subsection{Do-nothing}
\label{subsec:donothing}

\citet{JIN1993239} advect the outgoing waves that pass through the outlet
without reflecting them back into the domain for a mesh-based method using the
MOC. The equation which can be used to propagate a wave through the outlet is
given by
\begin{equation}
  \frac{\partial \ten{u}}{\partial t}+u \frac{\partial
  \ten{u}}{\partial x}- \nu \frac{\partial^{2} \ten{u}}{\partial
  y^{2}}=0,
  \label{eq:nrbcnew}
\end{equation}
where $\ten{u}$ is the velocity vector, $u$ is velocity component in $x$
direction and $\nu$ is the kinematic viscosity. In this paper, the diffusion
term has been dropped since the time for which outlet particles interact with
the fluid particles is not long enough for diffusion. Thus the
equation~\eqref{eq:nrbcnew} reduces to
\begin{equation}
  \frac{\partial \ten{u}}{\partial t}+u \frac{\partial
  \ten{u}}{\partial x}=0.
  \label{eq:nrbc2}
\end{equation}
\citet{ALVARADORODRIGUEZ2017177} proposes an SPH discretization of the
equation~\eqref{eq:nrbc2}, where the first term on the right-hand-side is
considered as a material derivative and the velocity is integrated by taking
the second term as acceleration. However, the equation~\eqref{eq:nrbc2}
physically means one must advect the particles in the normal direction to the
outlet while freezing all other properties like velocity and pressure. This is
similar to the method proposed by \citet{FEDERICO201235}. In SPH form, at the
outlet we can use
\begin{equation}
  x^{n}_{o} = x^{n-1}_{o} + u^{n-1}_{o}\Delta t,
  \label{}
\end{equation}
\begin{equation}
  u^{n}_o = u^{n-1}_{o}
  \label{}
\end{equation}
and
\begin{equation}
  p^{n}_{o} = p^{n-1}_{o},
  \label{}
\end{equation}
where $*^{n}_{o}$ denotes the outlet properties at time $n$ and $x$, $u$ and
$p$ are the position, $x$-component of the velocity and pressure respectively.

\subsubsection{Modified Do-nothing}
\label{subsubsec:mod_don}

We propose a subtle modification to the standard do-nothing method described
in section \ref{subsec:donothing}. Unlike the standard do-nothing where the
outlet moves with a velocity with which it left the fluid domain, we propose
to extrapolate the velocity of the fluid to the advection velocity of the
outlet particles.  Thus the advection is given by
\begin{equation}
	x^{n}_{o} = x^{n-1}_{o} + u^{n}_{ex}\Delta t,
  \label{}
\end{equation}
where $u^n_{ex}$ is the Shepard extrapolated fluid velocity at timestep $n$
given as
\begin{equation}
  \label{eq:shepard0}
  u_{ex} = \frac{\sum_j u_j W_{ij}}{\sum_j W_{ij}}.
\end{equation}
It must be noted that the advection velocities are only used to advect the
particles and the actual velocity of the outlet particles remain the ones
frozen when the fluid particle is converted to the outlet particle.

\subsection{Mirroring}
\label{subsec:mirror}

\begin{figure}[h!]
  \centering
  \resizebox{\linewidth}{!}{
    \begin{tikzpicture}[x=1cm, y=1cm, line width=0.1mm]
    \pgfmathsetmacro{\l}{20};
    \pgfmathsetmacro{\w}{10};

    \draw[-] (0, 0) -- (\l, 0);
    \draw[-] (\l, 0) -- (\l, \w);
    \draw[-] (\l, \w) -- (0, \w);
    \draw[-] (0, \w) -- (0, 0);

    \draw[-] (0.2*\l,0) -- (0.2*\l,\w);
    \draw[-] (0.8*\l,0) -- (0.8*\l,\w);

    \foreach \x in {1,...,20}
        \foreach \y in  {1,...,10}
            {
              \pgfmathsetmacro{\res}{ifthenelse(\x<5,"lightgray", ifthenelse(\x<17,"gray","darkgray")}
              \pgfmathsetmacro{\CosValue}{1/cos(\x)}
              \draw[fill=\res] (\x-0.5+\CosValue*0.1,\y-0.5-\CosValue*0.1) circle (0.3);
    };

    \foreach \x in {13,...,16}
        \foreach \y in  {1,...,10}
            {
              \pgfmathsetmacro{\CosValue}{1/cos(\x)};
              \draw[dashed, blue] (\x-0.73+\CosValue*0.1,\y-0.5-\CosValue*0.1) circle (0.3);
    };

    \foreach \x in {4,...,7}
        \foreach \y in  {1,...,10}
            {
              \pgfmathsetmacro{\CosValue}{1/cos(\x)};
              \draw[dashed, blue] (\x+0.3+\CosValue*0.1,\y-0.5-\CosValue*0.1) circle (0.3);
    };

    \node[yshift=10.2cm, xshift = 2.5cm] {Inlet};
    \node[yshift=10.2cm, xshift = 10cm] {Fluid};
    \node[yshift=10.2cm, xshift = 17.5cm] {Outlet};
    \draw (15.35, 5.40) circle(2.5);
    \draw[->] (15.35, 5.40) -- node[yshift=-0.3cm, xshift=0.0cm]{kh}(16.80, 7.5);
\end{tikzpicture}
  }
  \caption{Inlet outlet particle arrangement. The dashed blue circles
    represent the reflected particles of the inlet and outlet about the
    interface.}
  \label{fig:mirror_schematic}
\end{figure}
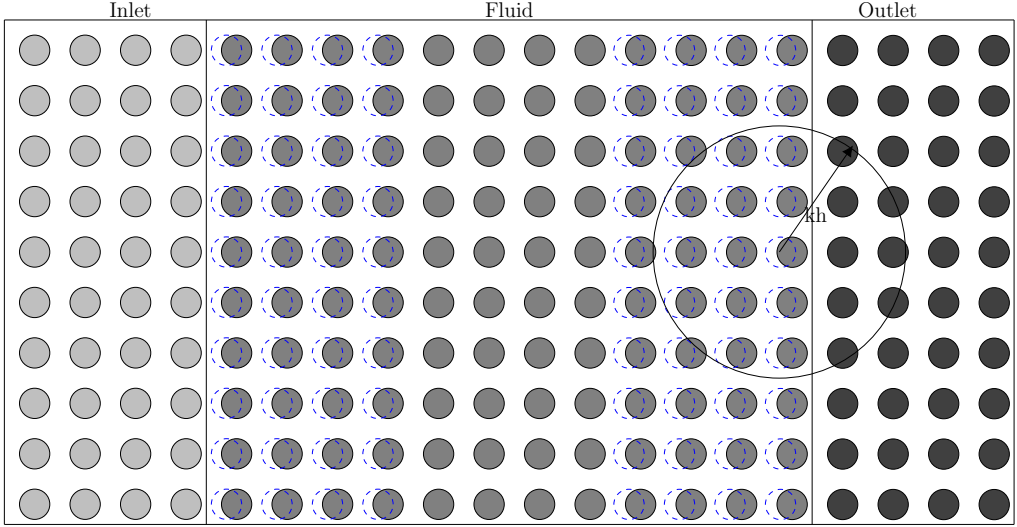

\citet{open_bc:tafuni:cmame:2018} employ a novel approach where the properties
at the inlet/outlet are extrapolated using a Taylor series expansion about a
mirrored particle at the fluid region. In the Fig.~\ref{fig:mirror_schematic},
we show the mirrored particles as circles with a dashed blue outline. The
mirror particles are generated by reflecting inlet/outlet particles about the
interface. Due to lack of kernel support at the interface, a higher order
approximation given by \citet{LIU200619} is used to determine the property
value at the mirrored particle.

In multiple dimensions, the first order Taylor series expansion for any
property $f$ of a particle about position $\mathbf{x}_k$ is
\begin{equation}
  f(\mathbf{x}) = f_{k} +f_{k,\beta}
  \left(\mathbf{x}-\mathbf{x}_{k}\right).
  \label{eq:liu1_taylor}
\end{equation}
Here $f_k = f(\ten{x}_k)$ and $f_{k,\beta}$ denotes the derivatives of the
function and $\beta \in {x, y, z}$. Taking the inner product of the function
with the SPH kernel $W_k(\ten{x}) = W(\ten{x} - \ten{x}_k)$ and it's
derivative $W_{k,\beta}(\ten{x}) = W_{\beta}(\ten{x} - \ten{x}_k)$, we obtain
\begin{equation}
  \int f(\mathbf{x}) W_{k} (\mathbf{x}) \mathrm{d} \mathbf{x} = f_{k} \int
  W_{k}(\mathbf{x}) \mathrm{d} \mathbf{x} + f_{k,\beta} \int
  \left(\mathbf{x}-\mathbf{x}_{k}\right) W_{k} (\mathbf{x}) \mathrm{d}
  \mathbf{x}
  \label{eq:liu1}
\end{equation}
and
\begin{equation}
  \int f(\mathbf{x}) W_{k,\beta} (\mathbf{x}) \mathrm{d} \mathbf{x} = f_{k}
  \int W_{k,\beta} (\mathbf{x}) \mathrm{d} \mathbf{x} + f_{k,\beta} \int
  \left(\mathbf{x} - \mathbf{x}_{k} \right) W_{k,\beta} (\mathbf{x})
  \mathrm{d} \mathbf{x}.
  \label{eq:liu2}
\end{equation}
Equations \eqref{eq:liu1} and \eqref{eq:liu2} can be written in matrix form
using SPH approximation as
\begin{equation}
  \left[\begin{array} {cccc} {W_{kl} V_{l}} & {x_{lk}  W_{kl} V_{l}} &
  {y_{lk} W_{kl} V_{l}} & {z_{lk} W_{kl} V_{l}} \\ {W_{kl,x} V_{l} } &
  {x_{lk}W_{kl,x} V_{l}} & {y_{lk}W_{kl,x} V_{l}} & {z_{lk}W_{kl,x} V_{l}}
  \\ { W_{kl,y} V_{l}} & {x_{lk}W_{kl,y} V_{l}} & {y_{lk}W_{kl,y}V_{l}} &
  {z_{lk}W_{kl,y}V_{l} } \\ {W_{kl,z}V_{l}} & {x_{lk}W_{kl,z}V_{l}} &
  {y_{lk}W_{kl,z}V_{l} } & {z_{lk}W_{kl,z}V_{l}} \end{array} \right] \left[
  \begin{array} {c} {f_{k}} \\ {f_{k,x}} \\ {f_{k,y}} \\ {f_{k,z}}
\end{array} \right] = \left[\begin{array} {c} {f_{l}W_{kl}V_{l} } \\
  {f_{l}W_{kl,x}V_{l}} \\ {f_{l}W_{kl,y}V_{l} } \\ {f_{l}W_{kl,z}V_{l} }
\end{array} \right],
  \label{eq:matform}
\end{equation}
where $W_{kl}$ and $W_{kl,\beta}$ : $\beta \in \{x,y,x\}$ are the kernel and
it's derivative respectively, $k$ denotes the destination index, $l$ denotes
the source particle index and $f$ is the property of interest, $V_k$ is the
volume of the $k$'th particle which in the present case is $m/\rho$, and
$x_{kl} = x_k - x_l$. For brevity repeated indices $l$, are summed over. Note
that the index $k$ indicates the target particle which is fixed and not summed
over. The above linear system is solved for each mirrored destination, which
gives the property and it's derivative. After evaluating $f$ and
$f_{k,\beta}$, the values at inlet/outlet are evaluated using the Taylor
series expansion given by
\begin{equation}
  f_{o} = f_{k} + \left(\mathbf{r}_{o} - \mathbf{r}_{k} \right) \cdot
  {\nabla} f_{k},
  \label{eq:taylor}
\end{equation}
about the corresponding ghost particle position $x_k$.
\citet{open_bc:tafuni:cmame:2018} extrapolate all the relevant properties
using equation \eqref{eq:taylor}. In this paper, we have modified the equation
for extrapolation to
\begin{equation}
  f_{o} = f_{k} - \left(\mathbf{r}_{o} - \mathbf{r}_{k} \right) \cdot
  {\nabla} f_{k}.
  \label{eq:taylor2}
\end{equation}
in order to get zero gradient at the outlet interface. When we use the
original form as written in \cite{open_bc:tafuni:cmame:2018}, the test cases
blow up.

\subsection{Method of characteristics}
\label{subsec:charac}

This method has been proposed by \citet{Lastiwka2009:nonrefbc}. The basic idea
is to resolve the perturbations from the mean flow in terms of the
characteristics and then use the characteristic variables to propagate the
appropriate values to the outlet or inlet. The scheme is itself based on the
work of \citet{giles:nrbc:aiaa:1990} who proposes general NRBCs for the Euler
equations.

The properties of fluid are rewritten in terms of the characteristic variables
perpendicular to the outlet. In this process the appropriate boundary
conditions may be applied. The following form of the characteristic variables
is used,
\begin{align}
  \nonumber
  J_1 &= -c_s^2 (\rho - \rho_{ref}) + (p - p_{ref}) \\ \label{eq:characteristic-vars}
  J_2 & = \rho c_s(u - u_{ref})  + (p - p_{ref}) \\   \nonumber
  J_3 & = -\rho c_s(u - u_{ref})  + (p - p_{ref}),
\end{align}
where the $u_{ref}, p_{ref}, \rho_{ref}$ denote the reference quantities in
the domain. We note that $J_1, J_2, J_3$ correspond to the quantities $c_1,
c_3, c_4$ in the work of \citet{giles:nrbc:aiaa:1990}. The outflow boundary
conditions basically require that $J_1$ and $J_2$ be determined from the
interior and that $J_3$ be set to zero. The perturbations in the plane of the
outlet pass through without any change, so the transverse components of the
velocity are not changed by this scheme.

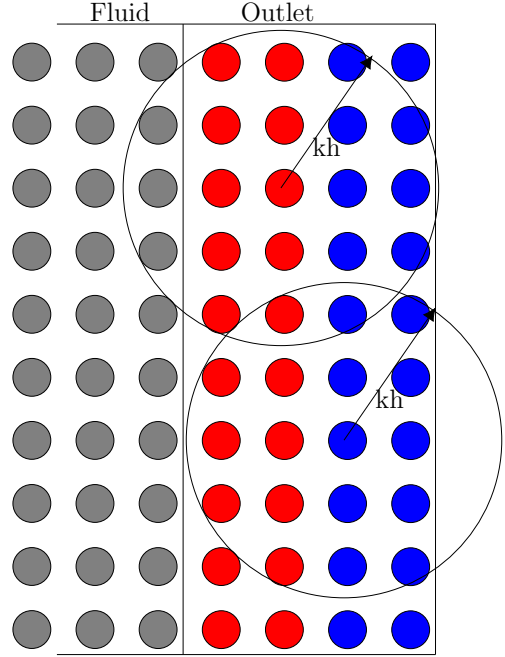
\begin{figure}[h!]
  \centering
  \resizebox{0.5\linewidth}{!}{
    \begin{tikzpicture}[x=1cm, y=1cm, line width=0.1mm]
    \pgfmathsetmacro{\l}{20};
    \pgfmathsetmacro{\w}{10};

    \draw[-] (0.7*\l, 0) -- (\l, 0);
    \draw[-] (\l, 0) -- (\l, \w);
    \draw[-] (\l, \w) -- (0.7*\l, \w);

    \draw[-] (0.8*\l,0) -- (0.8*\l,\w);

    \foreach \x in {14,...,20}
        \foreach \y in  {1,...,10}
            {
              \pgfmathsetmacro{\res}{ifthenelse(\x<17,"gray", ifthenelse(\x<19,"red","blue")}
              \pgfmathsetmacro{\CosValue}{1/cos(\x)}
              \draw[fill=\res] (\x-0.5+\CosValue*0.1,\y-0.5-\CosValue*0.1) circle (0.3);
    };

    \node[yshift=10.2cm, xshift = 15cm] {Fluid};
    \node[yshift=10.2cm, xshift = 17.5cm] {Outlet};
    \draw (18.55, 3.40) circle(2.5);
    \draw (17.55, 7.40) circle(2.5);
    \draw[->] (18.55, 3.40) -- node[yshift=-0.4cm, xshift=0.0cm]{kh}(20.00, 5.5);
    \draw[->] (17.55, 7.40) -- node[yshift=-0.4cm, xshift=0.0cm]{kh}(19.00, 9.5);
\end{tikzpicture}
  }
  \caption{The outlet particles having fluid particles in their support radius
    are shown in red and without fluid particles in their support are in blue.}
  \label{fig:io_moc}
\end{figure}

In our implementation we use a simple Shepard interpolation given by
\begin{equation}
  \label{eq:shepard}
  f_{i} = \frac{\sum_{j}^{N} f_{j} W_{ij} }{\sum_{j}^{N} W_{ij}},
\end{equation}
where the $i^{th}$ outlet particle is outside the fluid domain as shown in
Fig.~\ref{fig:io_schematic} and $f$ is either $J_1$ or $J_2$. We use
equation~\eqref{eq:shepard} to interpolate $J_1, J_2$ from the fluid to the
outlet. Note that only some of the outlet particles are in the influence of
the fluid. Fig.~\ref{fig:io_moc} shows a sketch of the outlet and fluid. As can be seen, the
red particles are under the influence of the fluid but the blue particles are
not. For the blue particles in the outlet which are outside the influence of
the fluid particles i.e.\ $N=0$, we find the $J_1$ and $J_2$ using the average
of the values of red particles but at the previous timestep,
\begin{equation}
  \label{eq:average_old}
  f_{i}^{n} = \frac{\sum_{j}^{M} f_{j}^{n-1}}{M},
\end{equation}
where $M$ are the number of red particles which are in the support of the
$i$\textsuperscript{th} blue particle. Given $J_1, J_2, J_3$ we can easily
solve for the actual variables $u, p, \rho$ using
equation~\eqref{eq:characteristic-vars}.

\subsection{A new hybrid method}
\label{subsec:hybrid}
In this section, we describe a new method to implement outlet boundaries. At
the outlet, essentially two kinds of fluctuations are encountered namely
spatial variations which do not change rapidly in time and variations due to
acoustic waves which travel with the speed of sound. The weakly compressible
SPH schemes generate perturbations that travel with the prescribed speed of
sound unlike with ISPH schemes which solve for a pressure-Poisson equation. In
the case of a do-nothing type of outlet boundary as described earlier, the
particle properties are frozen. As a result, when the acoustic wave arrives at
the outlet, its velocity suddenly drops to the particle velocity in the
outlet. This causes an increase in the pressure for particles that are near
the outlet. In our proposed method, we devise a method to separate the fluid
flow properties into acoustic and base flow properties.

A time averaged property of the flow is given by
\begin{equation}
  f_{avg} = \frac{\sum_{n=1}^{N} f_n}{N},
  \label{eq:timeevg}
\end{equation}
where $f$ is the fluid property, $N$ is the number of time steps used in the
averaging. The value of $N$ can be estimated by determining the number of time
steps the acoustic wave takes to move from one particle to another given by
\begin{equation}
	N = \frac{\Delta x}{\Delta t (u+c_s)}.
	\label{eq:N}
\end{equation}
In all our cases $N \approx 4$, thus in order to have a sufficient time
average we take $N = 6$ for all our test cases. Further, in order to detect
the acoustic wave, the acoustic intensity is used as a parameter. The time
averaged properties are not changed whenever the acoustic intensity of the
flow is greater than the prescribed value. The acoustic intensity is given by
$p^2/(2\rho c_s)$ \cite{kinsler1999fundamentals}. The prescribed value of
acoustic intensity can be determined using the inlet velocity, $u_i$ and is given by
\begin{equation}
  I = \frac{(\frac{1}{2} \rho u_i^2)^2}{2 \rho c_s}.
  \label{eq:acouinten}
\end{equation}
The difference between the particle property and its time-average gives us the
acoustic component. The time-averaged part is advected out of the domain using
the do-nothing method. Since the acoustic wave travels with the speed of sound
it should be propagated out with the same. We use the method of
characteristics described in the previous section to propagate these acoustic
perturbations into the outlet, where the reference values are the
time-averages. In our implementation, we keep $\rho_{ref}$ fixed.

When a particle moves from the fluid domain into the outlet, it retains its
time average values. The acoustic properties are added to this using Shepard
interpolation to the outlet zone as
\begin{equation}
	f_o = f_{ac} + f_{avg},
  \label{eq:acou}
\end{equation}
where $f_{ac}$ is determined using the extrapolated $J_2$ as explained in
section \ref{subsec:charac}. Since the do-nothing condition is used at the
outlet for the time-averaged values, the proposed method cannot simulate
incoming flow near the outlet however it is suitable for wind tunnel type of
flow where the flow always exits the outlet from one side. The particles in
the outlet layer are advected using the velocity evaluated with equation
\eqref{eq:acou} (assuming the outlet is perpendicular to the x-axis). The
particles are not moved in the transverse direction. We note that the Shepard
interpolation of the properties from the fluid will not always carry to all
the particles in the outlet. These particles are advected with the average of
the existing outlet advection velocity.

For all the inlet/outlet methods described here, inlet particles are added to
the fluid particles whenever they cross the inlet-fluid interface. Similarly,
at the outlet, fluid particles are removed and added to outlet whenever they
cross the fluid-outlet interface. The particles are deleted once they leave the
outlet region.

\section{Results and discussion}
\label{sec:results}

In this section, we compare the different methods for the outlet boundary
condition with a variety of test cases. Each of these cases only takes a small
amount of computational effort and highlights specific issues. The new
problems are all two-dimensional and this makes them relatively easy to
implement. They include a one dimensional pulse (in a two-dimensional domain),
a two-dimensional pulse, a two-dimensional vortex, and a ramp inlet condition
in order to test the typical conditions that outlets encounter. In order to
obtain a solution representing an infinite domain for comparison, we simulate
the flow in a very long domain. The properties of the fluid are measured at a
probe placed inside the domain at a distance $d$ from the inlet. The length of
the domain, $L$ is chosen to be $d + c_st$, and $t$ is the simulation time. We
treat the fluid as inviscid and use a particle spacing of $\Delta x = 0.1$
unless stated otherwise in all our testcases. We use the results in the long
domain as a reference and use this to compute the $L_{2}$ norm of the errors
in the various properties using
\begin{equation}
  \label{eq:l2norm}
  e(f) = \left( \frac{\sum_n (f^n - f^n_l)^{2}}{\sum_n (f^{n}_l)^{2}} \right)^{1/2},
\end{equation}
where $n$ represents the timestep, $f^n$ is the property of interest at a
particular timestep and $f^n_l$ is the corresponding property in the long
domain. Once these test cases are simulated we demonstrate the best of these
methods for an impulsively started flow past a circular cylinder at different
Reynolds numbers and a backward-facing step.

\subsection{1D Pressure bump}

\begin{figure}[h!]
  \centering
  \includegraphics[scale=0.7]{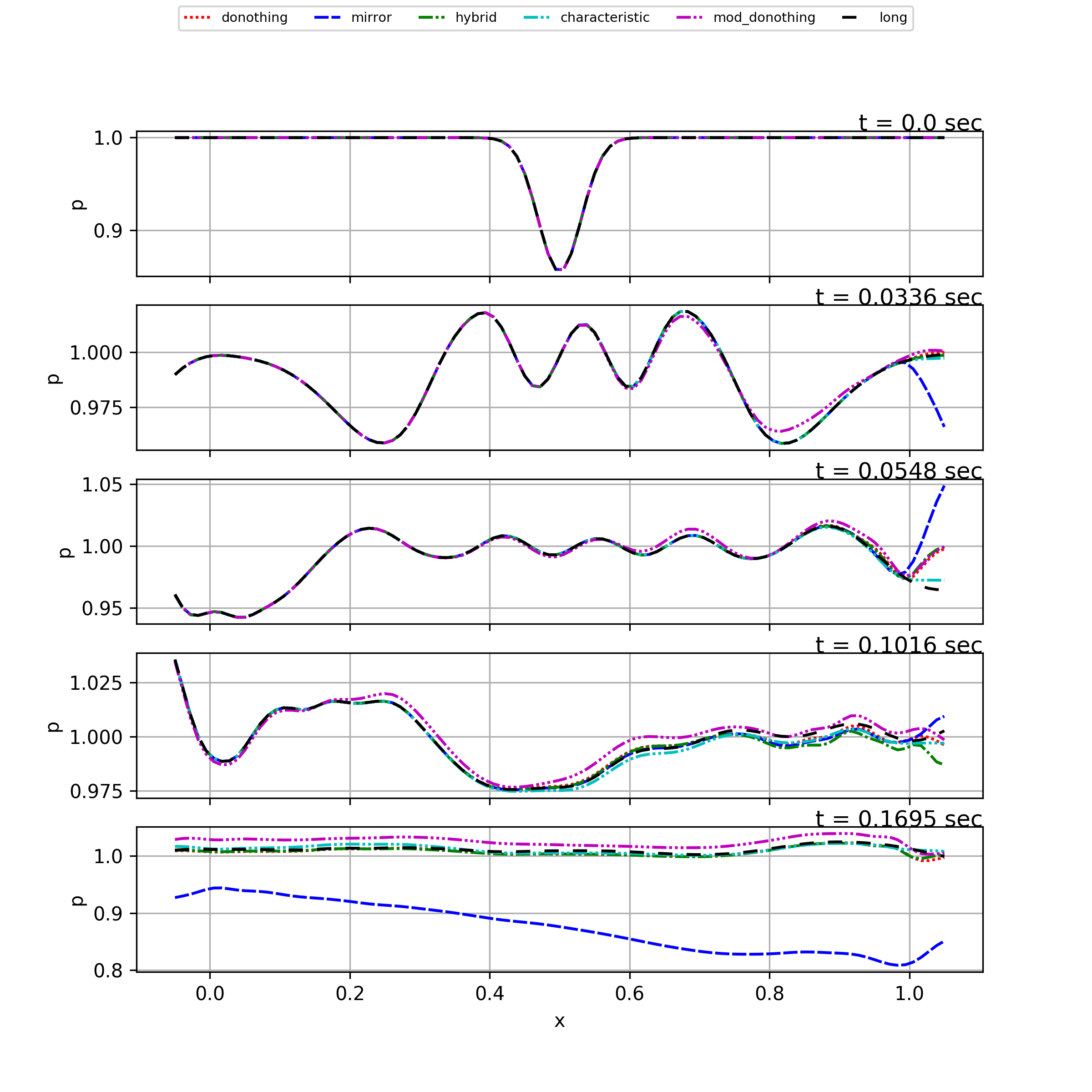}
  \caption{Pressure plot at various times for the different methods.  The
    solid line denotes the solution with the long domain.}
  \label{fig:pressurepulse}
\end{figure}
This test case was proposed by \citet{Lastiwka2009:nonrefbc}. In this testcase,
the fluid domain is initialized with a pressure variation given by
\begin{equation}
	p(x) = 1.0 - 0.2 e^{\frac{-(x-0.5)^2}{0.001}}.
	\label{eq:pressurebump}
\end{equation}
The pressure at inlet and outlet is initialized with $p=1.0$. Velocity of the
domain including inlet and outlet remain constant (=1m/s) for all times. The
domain length is $1m$ and the pressure bump is at $x=0.5m$. We use the
artificial viscosity parameter, $\alpha=0.1$ as mentioned in
\cite{Lastiwka2009:nonrefbc}. We simulated the testcase for all the types of
outlet boundaries described in the section \ref{sec:bc}. In case of the MOC
for all the test cases $u_{ref}, p_{ref} $ and $\rho_{ref}$ is taken as $1.0
m/s$, $1.0\ Pa$ and $1000 kg/m^3$ respectively. In
Fig.~\ref{fig:pressurepulse}, we compare the pressure along the centerline of
the domain at different times for all the methods. It can be seen that
mirroring technique results in a significant drop in pressure towards the end.
The modified do-nothing increases the pressure in the domain by a small
amount. All other cases, match well with the MOC and with the long domain.

\subsection{2D pulse}
\label {subsec:2dpulse}

\begin{figure}[h!]
	\centering
  \includegraphics[width=0.45\textwidth]{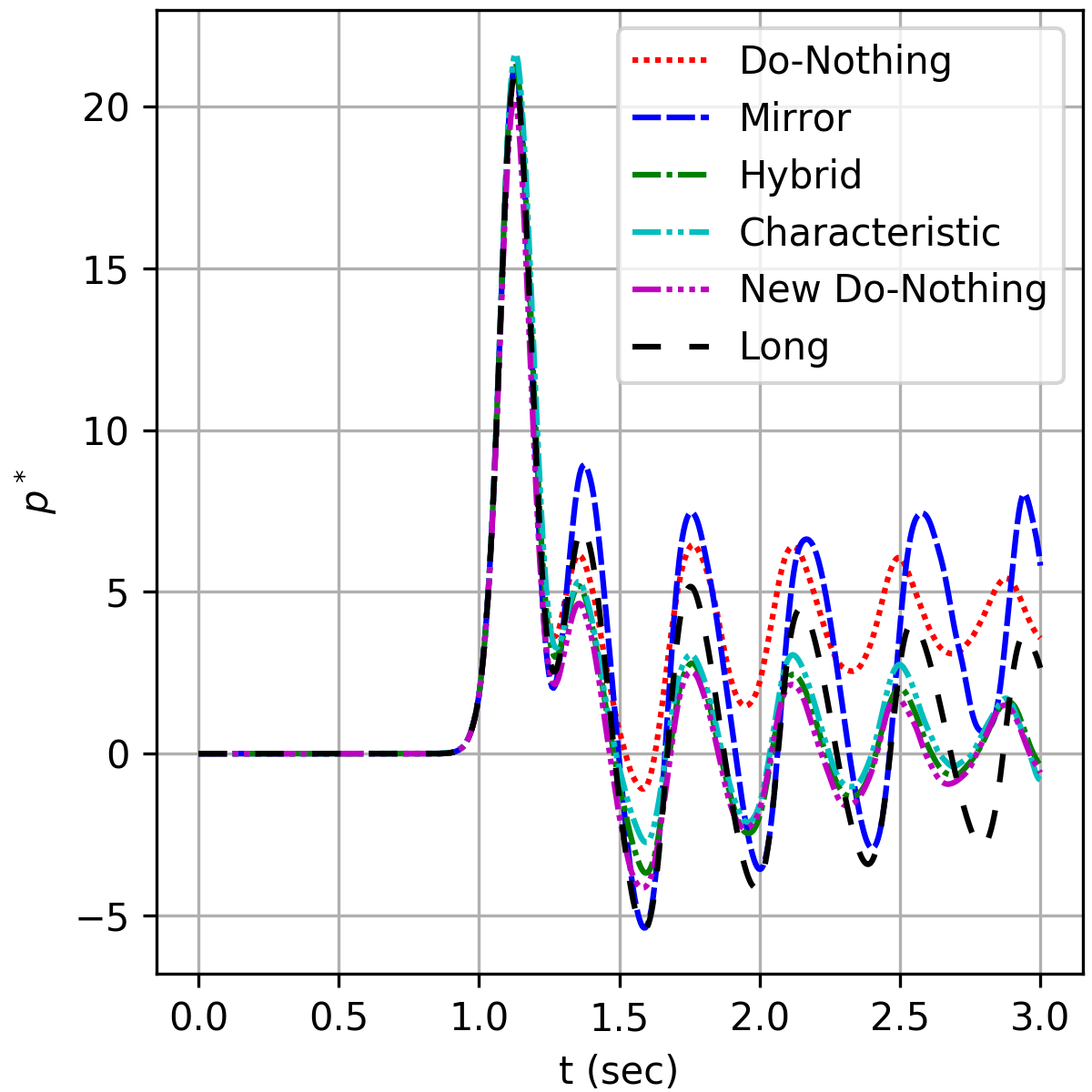}
  \includegraphics[width=0.45\textwidth]{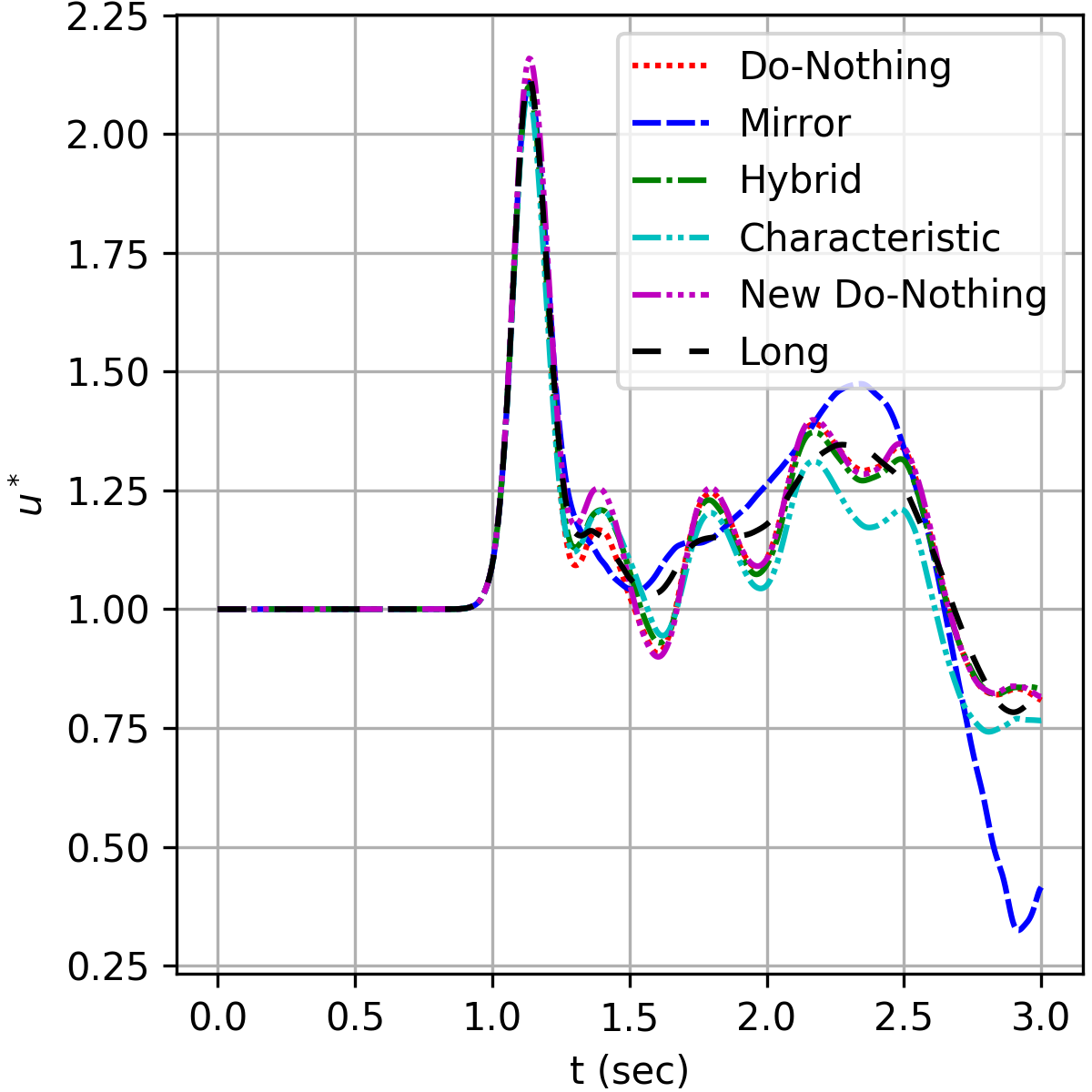}
	\caption{Normalized pressure (left) and velocity (right) plots at
    $x=1.7m$ with time for 2D varying inlet.}
	\label{fig:pulse}
\end{figure}

This benchmark tests the non-reflectivity for a two-dimensional disturbance. A
2D domain is considered, consisting of fluid with domain length, $L=2 m$ and
width, $W=2 m$. The probe is placed at $d=1.7m$ from the inlet. The fluid
region is constrained by inviscid walls on both sides. The inflow is taken
from the left and outlet is kept at the right of the fluid. The inlet, wall,
and outlet are initialized with $6$ layers of particles. In order to introduce
a 2D variation, the $u$ velocity is made a function of $y$, given by
\begin{equation}
  u(x,y,t) =  \left\{ \begin{array} { l l } {1.0 + 0.5 \cos \left(
        \frac{ \pi y}{12} \right)
      e^{\frac{{(t-1)}^{2}}{\delta} }} & {
			1.0 < t < 1.1} \\ {1.0
      } & { \text{elsewhere} }
  \end{array} \right..
  \label{eq:velfunc2}
\end{equation}
We normalize $p$ and $u$ measured at the probe such that $u^*=u/u_{ref}$ and
$p^*=\frac{2p}{\rho u_{ref}^2}$ respectively. Fig.~\ref{fig:pulse} shows the
plot of $u^*$ and $p^*$ versus time for the different outlets and
Table~\ref{table:pulse2d_l2} shows $L_2$ errors in the pressure and velocity
for the different outlet implementations.

The pressure variation with the MOC and hybrid methods are very close to the
results for a long domain compared to mirroring and do-nothing outlet. It can
be seen that the mirroring technique generates a lot of reflections into the
fluid as compared to do-nothing and MOC. In case of the do-nothing a
significant increase in pressure can be seen just after the wave passes
through the outlet (at around $1.25s$).

\begin{table}[h!]
\centering
\begin{tabular}{lrr}
\toprule
        Methods & $e(p^{*})$ & $e(u^{*})$ \\
\midrule
 Characteristic &      0.328 &      0.057 \\
     Do-Nothing &      0.629 &      0.038 \\
         Hybrid &      0.311 &      0.035 \\
         Mirror &      0.409 &      0.106 \\
 New Do-Nothing &      0.341 &      0.042 \\
\bottomrule
\end{tabular}

\caption{$L_2$ error in the $p^*$ and $u^*$ measured at the probe for the 2D pulse
problem.}
\label{table:pulse2d_l2}
\end{table}
Looking at the variation of the velocity we can see that both the modified
do-nothing and the new hybrid method show a close match to the results for a
long domain. After $2s$, the MOC method differs from the long-domain results
due to the spatial variations arriving near the outlet. The modified
do-nothing method is clearly better than the standard do-nothing scheme. These
conclusions are also borne out by the values of the $L_2$ norm as seen in
Table~\ref{table:pulse2d_l2}. The proposed hybrid method has the least errors.
\begin{figure}[h!]
	\centering
  \includegraphics[width=0.45\textwidth]{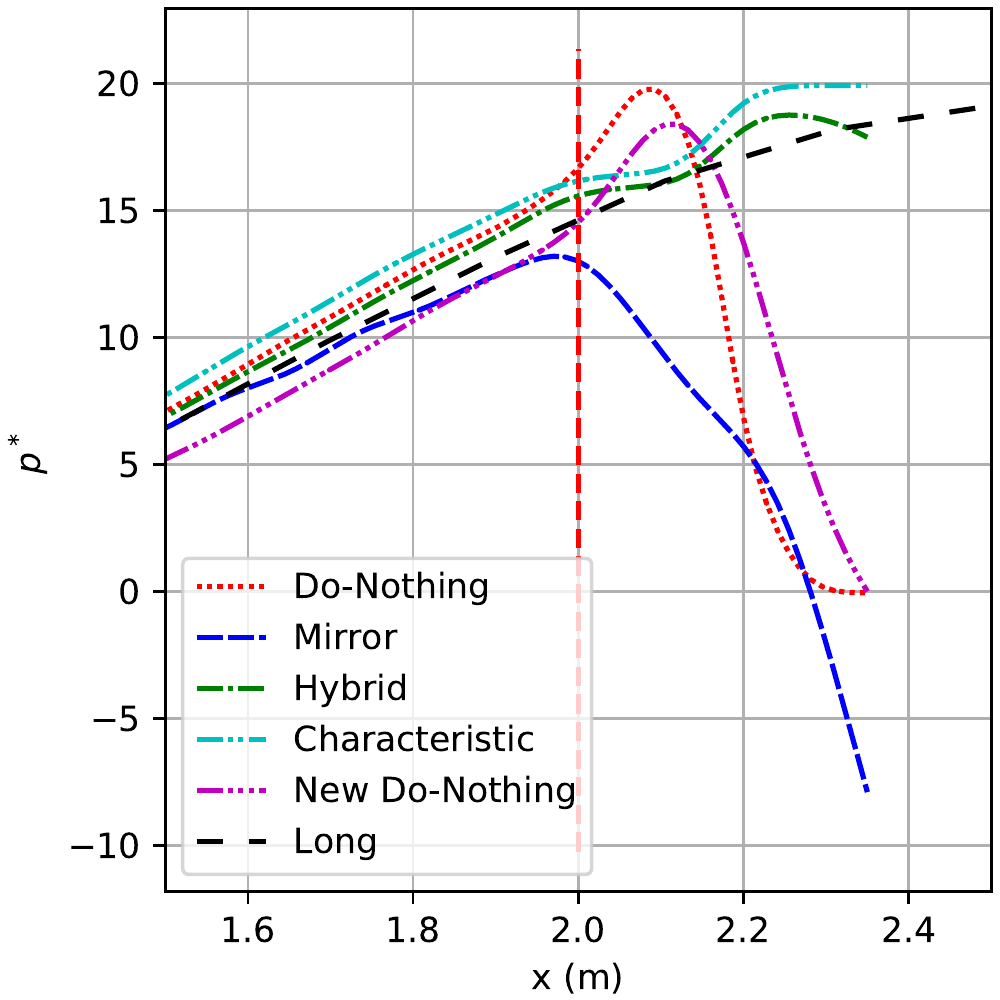}
  \includegraphics[width=0.45\textwidth]{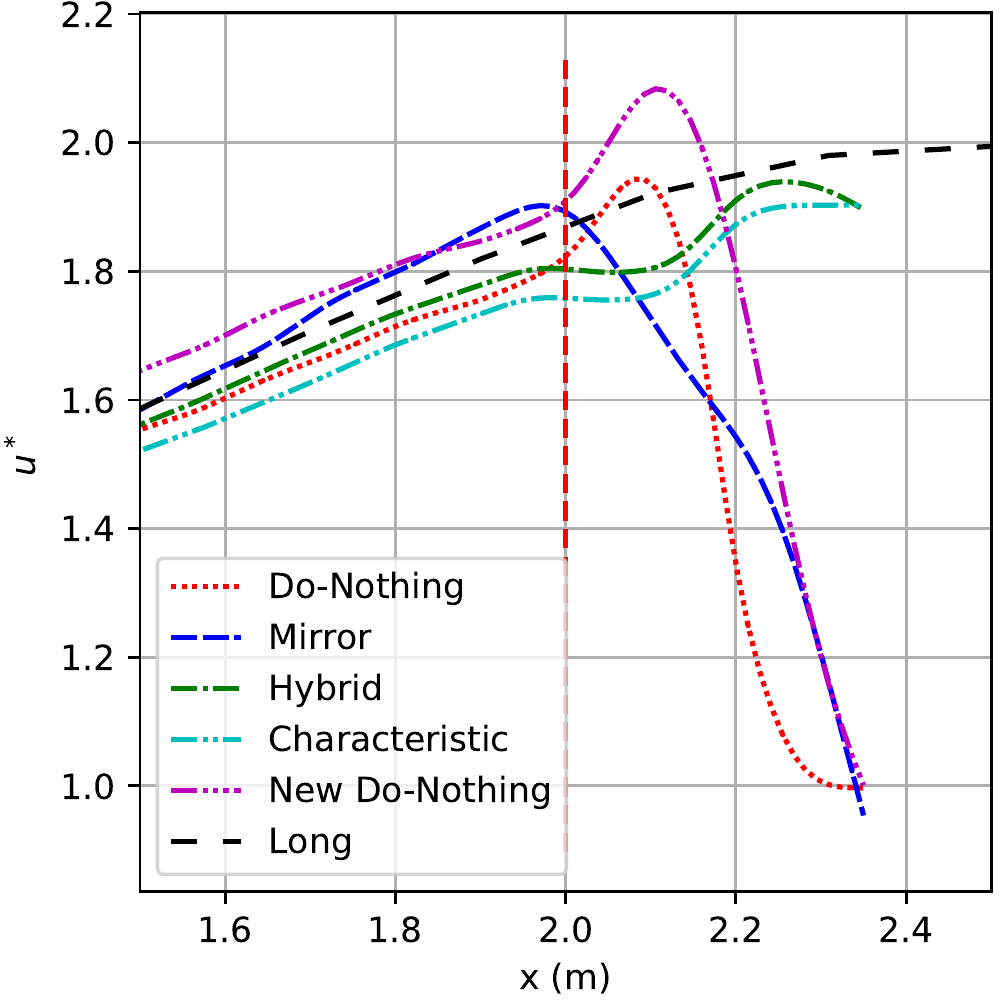}
  \caption{Normalized pressure (left) and velocity (right) along the $y=0$ line for
    the 2D pulse problem. Left of the dashed red line is fluid and right is
    outlet region.}
	\label{fig:extrapolation}
\end{figure}

\begin{figure}[h!]
	\centering
  \includegraphics[width=0.45\textwidth]{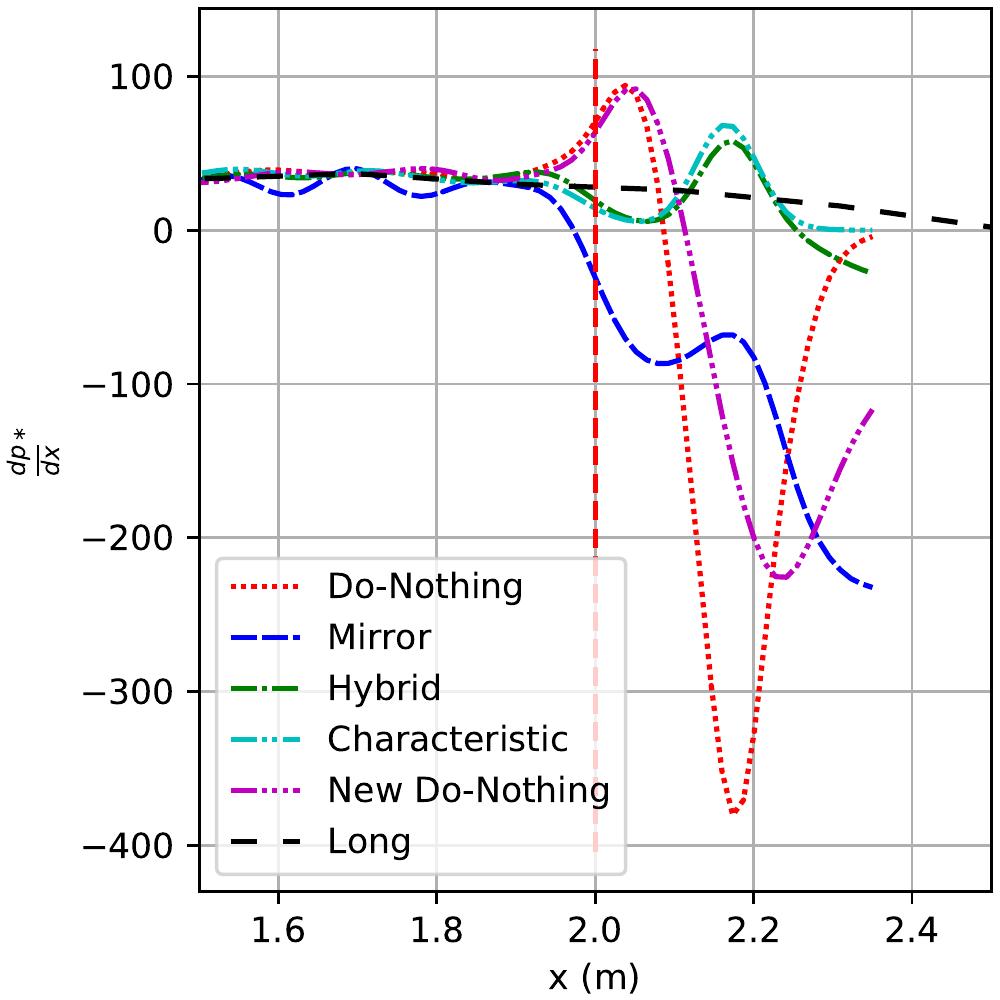}
  \includegraphics[width=0.45\textwidth]{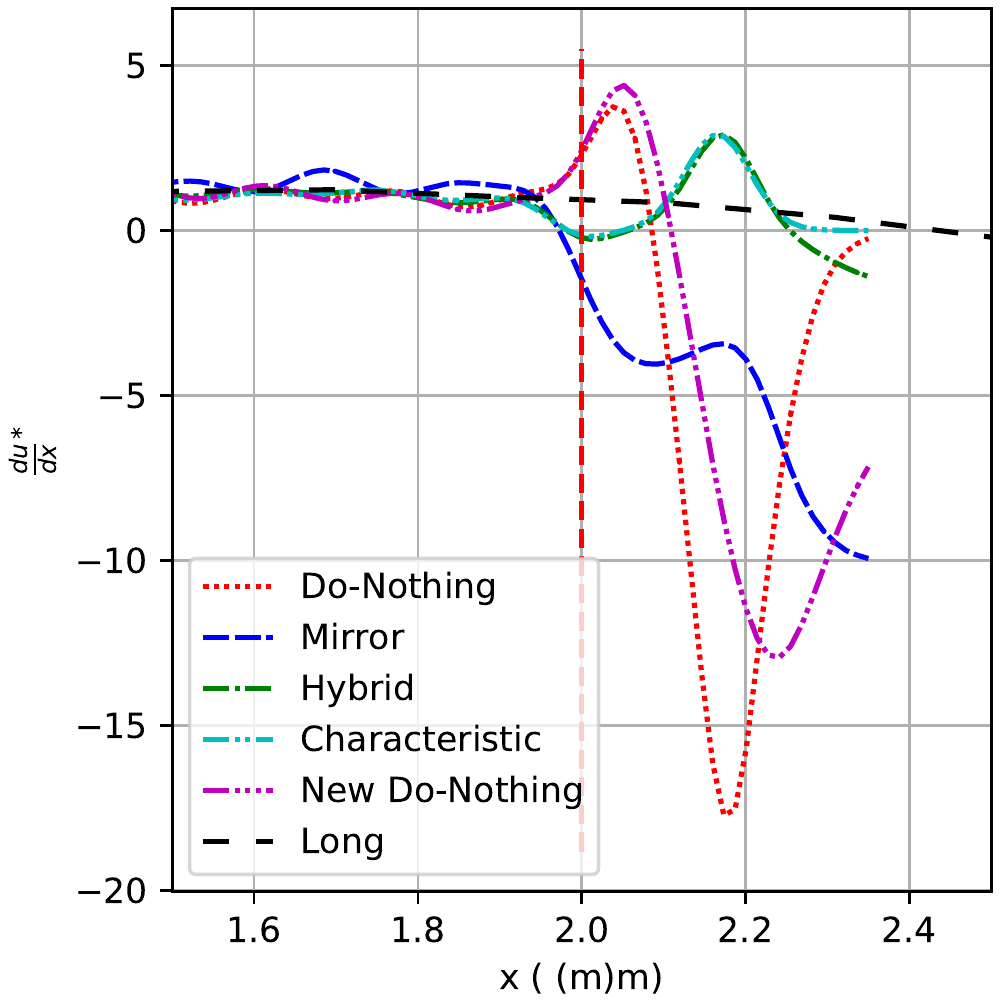}
  \caption{Normalized pressure (left) and velocity gradients (right) along the
    $y=0$ line for the 2D pulse problem. Left of the dashed red line is fluid
    and right is outlet region.}
	\label{fig:extrapolationgrad}
\end{figure}
In order to show the nature of the property variation across the fluid outlet
interface due to extrapolation, we interpolated pressure, velocity and their
gradients on a $y=0$ line as shown in Figure \ref{fig:extrapolation} and
\ref{fig:extrapolationgrad}. It can be seen that, in case of the mirroring
technique that the gradient of the property is zero at the interface. The
property is mirrored about the domain boundary however both hybrid and
do-nothing retain the history of the particle such that velocity and pressure
in the outlet do not affect the upstream flow. On looking at the gradient
along $x$ of the property for all the methods in
Fig.\ref{fig:extrapolationgrad}, we find that the mirroring technique impose
natural boundary conditions on fluid particles near the outlet i.e $\partial u/
\partial x = 0$, and $\partial p /\partial x =0$. In case of do-nothing and
modified do-nothing, the velocity and pressure profiles matched the long
domain but gradient changes significantly. However, the method of
characteristics and hybrid maintains the flow gradients along with the flow
variables as they are. In the context of the SPH, the latter seems to be very
important.

As discussed in section~\ref{sec:sph}, the EDAC method involves a parameter
called $\alpha$ which increases the pressure damping. We explore varying the
parameter $\alpha$ and study the error in $p^*$ for the different schemes in
Table~\ref{table:pulse2d_edac_p}. It can be observed that as $\alpha$
increases the pressure oscillations are reduced and therefore the errors
reduce for all the schemes. However, the greatest reduction is for the
original do-nothing and mirror methods. The others are not significantly
affected. This suggests that the hybrid method and modified do-nothing are
robust techniques.

\begin{table}[h!]
\centering
\begin{tabular}{lrrrr}
\toprule
        Methods & $\alpha = 0.1$ & $\alpha = 0.2$ & $\alpha = 0.5$ & $\alpha = 1.0$ \\
\midrule
 Characteristic &          0.336 &          0.334 &          0.328 &          0.315 \\
     Do-Nothing &          1.135 &          0.864 &          0.629 &          0.587 \\
         Hybrid &          0.339 &          0.319 &          0.311 &          0.306 \\
         Mirror &          0.533 &          0.463 &          0.409 &          0.371 \\
 New Do-Nothing &          0.391 &          0.543 &          0.341 &          0.354 \\
\bottomrule
\end{tabular}

\caption{$L_2$ error in $p^*$ measured at the probe for the 2D pulse as the
  EDAC parameter $\alpha$.}
\label{table:pulse2d_edac_p}
\end{table}

\subsection{1D ramp}
\label{subsec:ramp}

In this test case, we impose a ramp velocity on the inlet particles such that
$u =0 m /s$ at $t = 0 s$ and $u = 1 m/s$ at $t = 1 s$. After time $t = 1s$,
the velocity is fixed at $1 m/s$. The size of domain, boundary condition and
initialization are same as in the case of the 2D pulse. We simulate the test
case for each method and compare it with results for a long domain.
\begin{figure}[h!]
  \centering
  \includegraphics[width=0.45\textwidth]{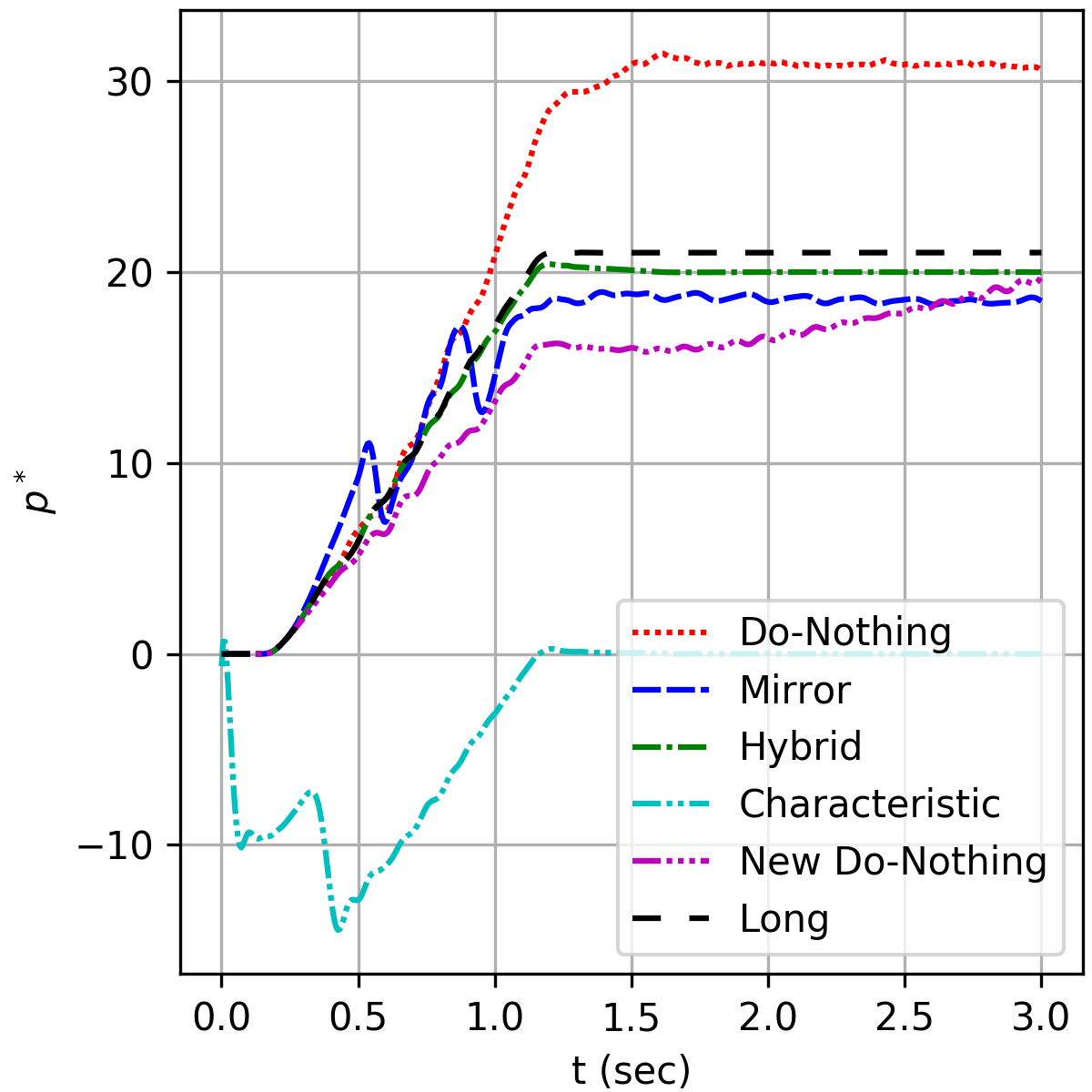}
  \includegraphics[width=0.45\textwidth]{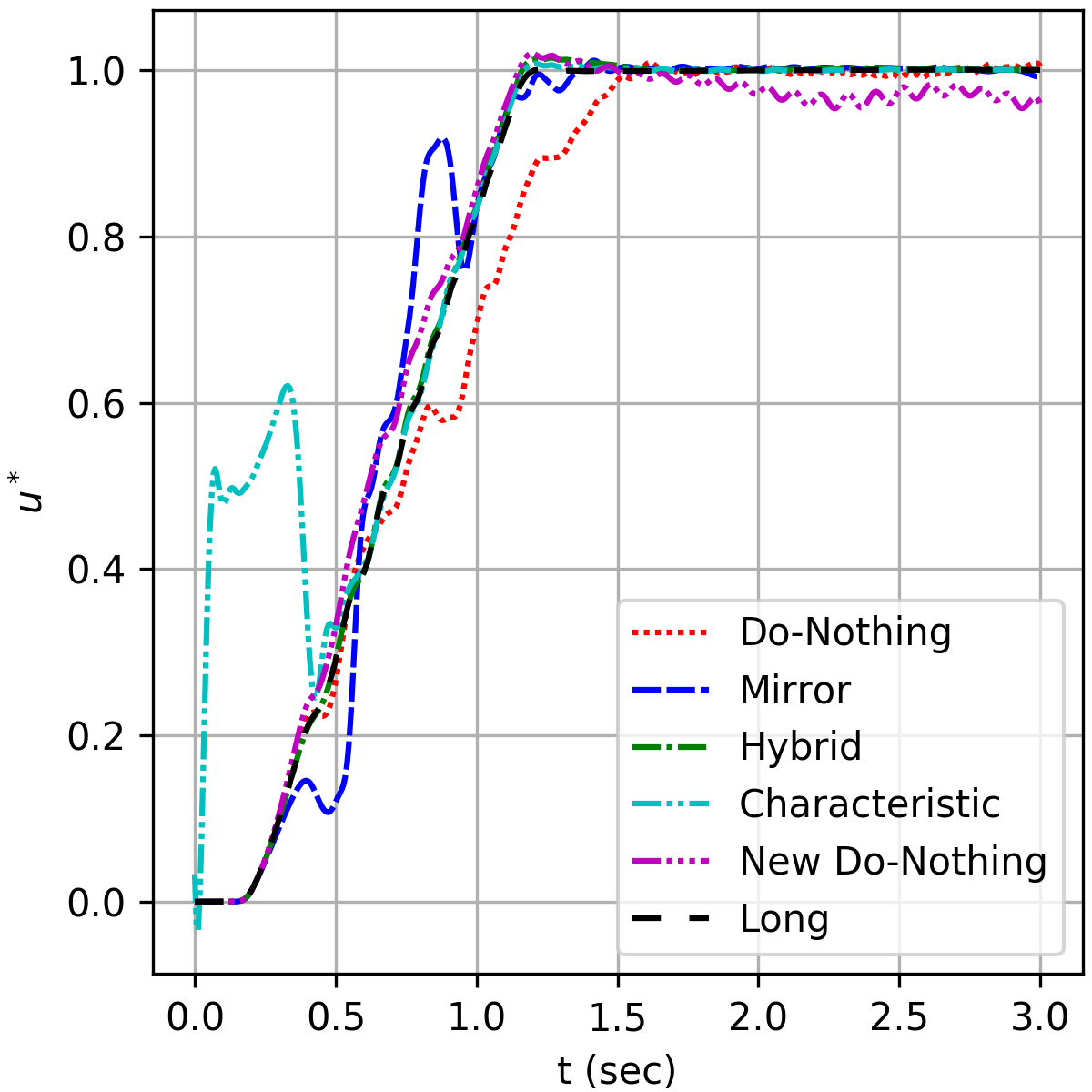}
  \caption{Normalized pressure (left) and velocity (right) plots at
  $x=1.7m$ with time for ramp inlet.}
  \label{fig:ramp_pu}
\end{figure}
In the Figure \ref{fig:ramp_pu}, we have plotted the $p^*, u^*$ for the ramp
and the $L_2$ errors are shown in Table~\ref{table:ramp_l2}. In case of the
pressure, the hybrid, mirror, and modified do-nothing methods work well. The
standard do-nothing method generates a significantly high pressure as the
initial particles at the outlet do not move and thereby cause an increase in
pressure. In case of the MOC, there is no specific method to determine the
reference values for $u, p$ at the initial stage and this seems to cause the
problems. Similar issues are seen in the case of the velocity for the MOC. As
seen in Table~\ref{table:ramp_l2}, the hybrid method has the least errors for
both pressure and velocity.

\begin{table}[h!]
\centering
\begin{tabular}{lrr}
\toprule
        Methods & $e(p^{*})$ & $e(u^{*})$ \\
\midrule
 Characteristic &      1.099 &      0.193 \\
     Do-Nothing &      0.433 &      0.066 \\
         Hybrid &      0.043 &      0.007 \\
         Mirror &      0.123 &      0.074 \\
 New Do-Nothing &      0.197 &      0.039 \\
\bottomrule
\end{tabular}

\caption{$L_2$ error in the $p^*$ and $u^*$ measured at the probe for the ramp
  velocity problem.}
\label{table:ramp_l2}
\end{table}

\subsection{2D vortex}
\begin{figure}[h!]
  \centering
  \includegraphics[width=0.45\textwidth]{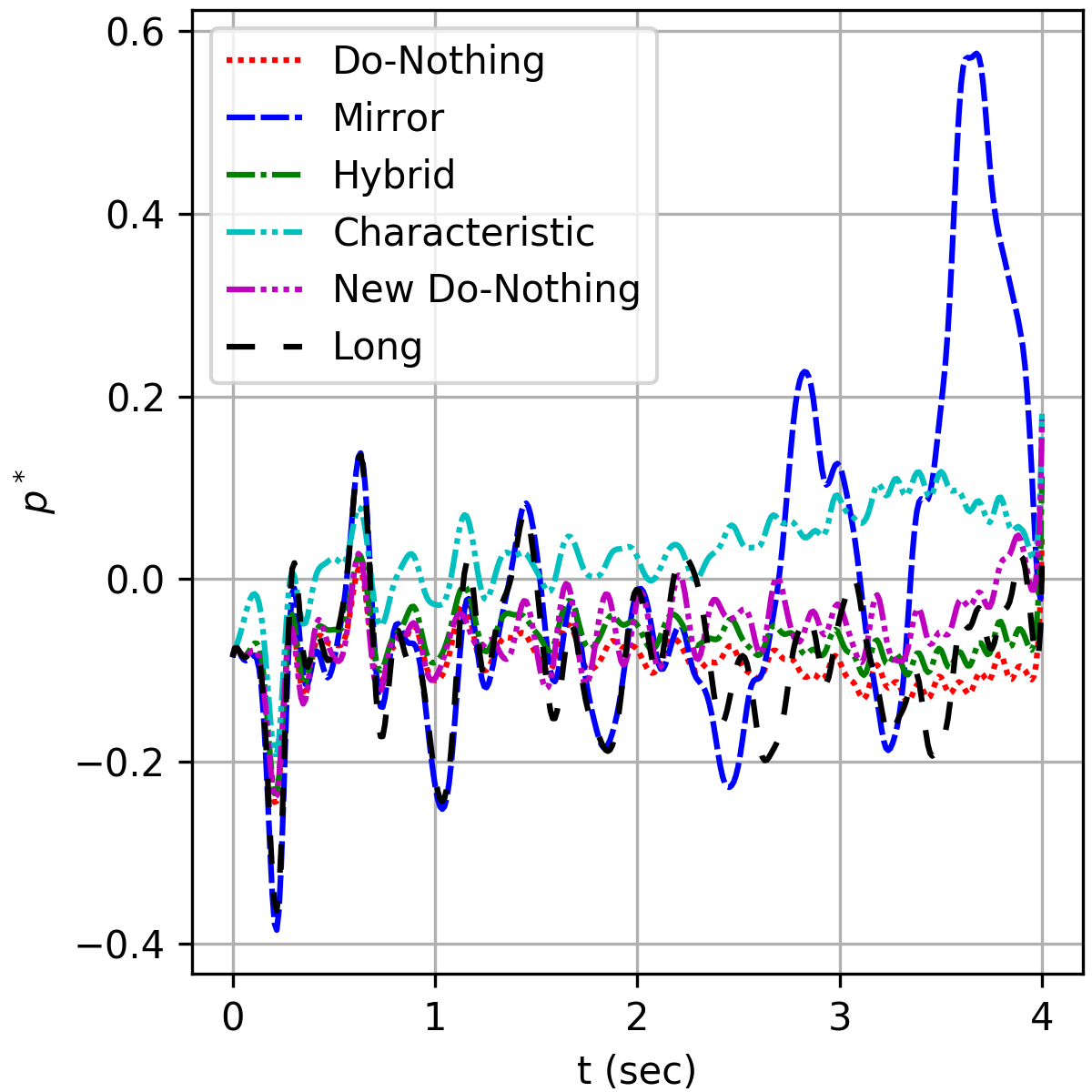}
  \includegraphics[width=0.45\textwidth]{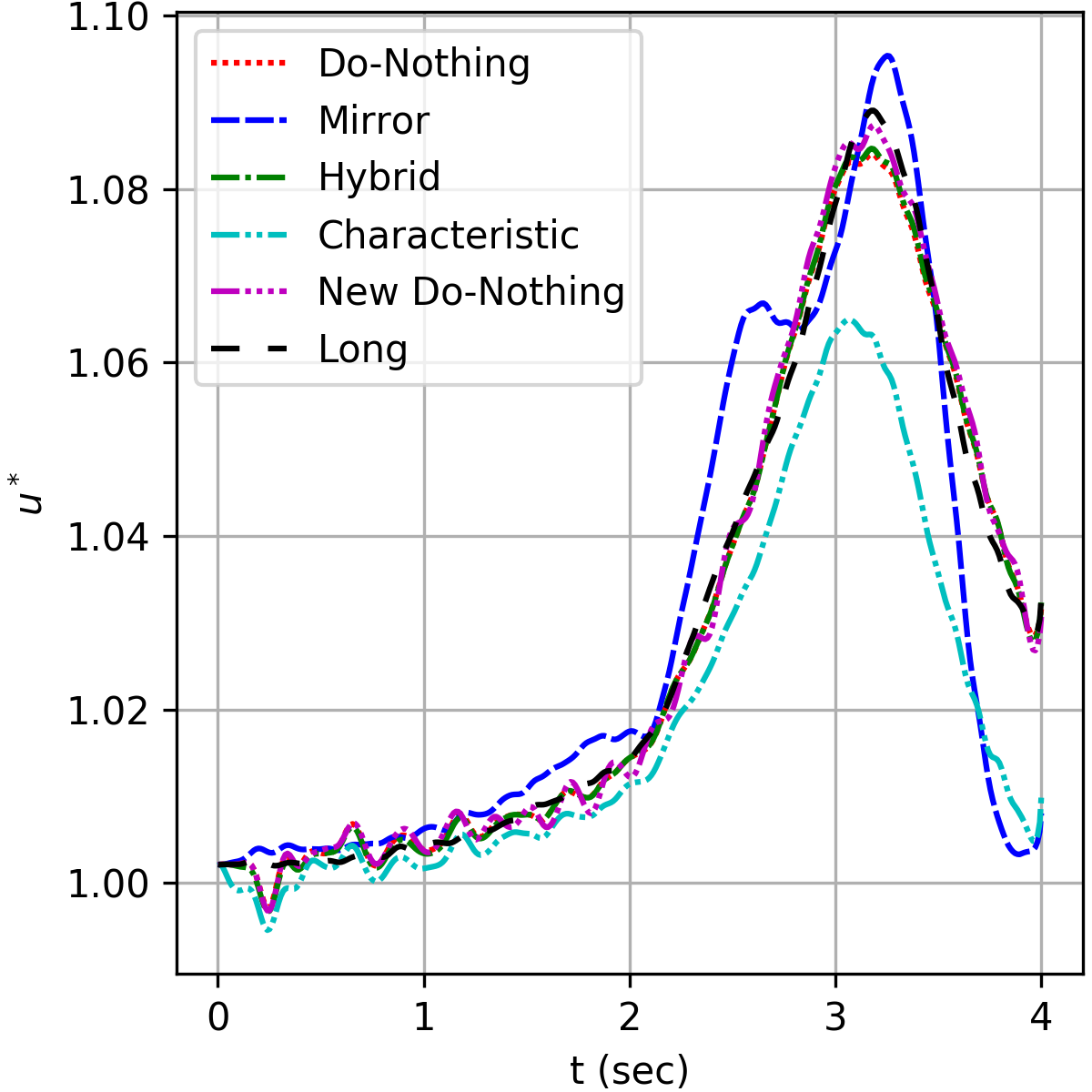}
  \includegraphics[width=0.45\textwidth]{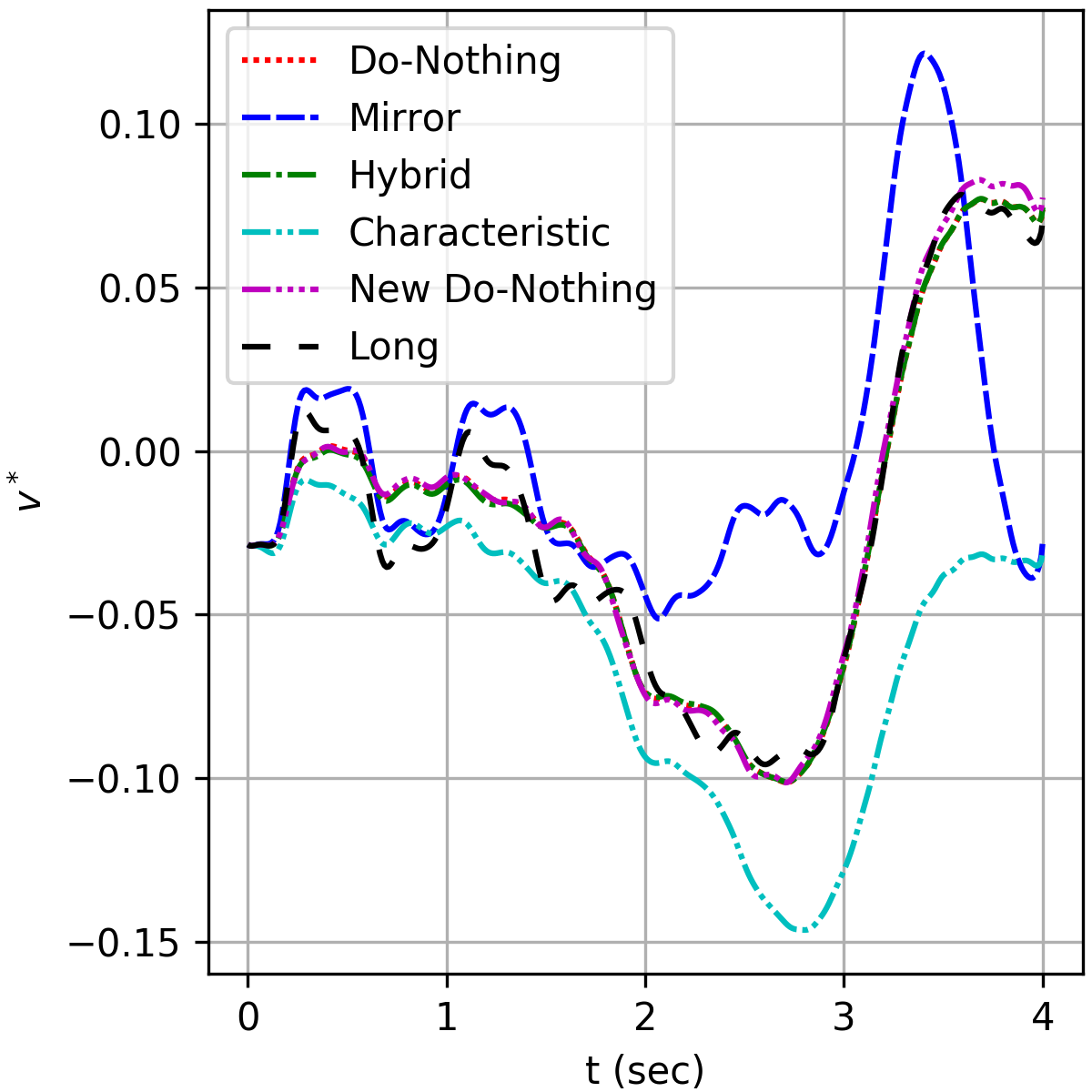}
  \caption{Normalized pressure (left), u-velocity (right) and v-velocity
    (center) plots at $x=1.7m$ with time for 2D vortex advection with $1m/s$.}
  \label{fig:vortex_pu}
\end{figure}
In this test case, a vortex is generated in the inlet moving with a constant
velocity of $1m/s$ and allowed to go through the outlet. This case tests the
outlet for permeability for a velocity variation similar to vortex shedding.
It is important that this be preserved for most engineering flow simulations.
The domain size is kept same as in case of 2D pulse however the width is
doubled to accommodate the vortex. The vortex is generated by changing the
velocity at the inlet with time using
\begin{equation}
  (u, v) = \left(1.0 + \frac{\Gamma y}{r^2 + 0.2}, \frac{-\Gamma x}{r^2 +
  0.2}\right),
  \label{eq:vortex}
\end{equation}
where $\Gamma=0.1$ is the vortex strength, and $r = \sqrt{x^2+y^2}$ is the
distance from the center of the vortex. In order to calculate the distance of
the vortex we used
\begin{equation}
  x = u (1 - t),
  \label{eq:vortexx}
\end{equation}
where $u$ is the speed of the vortex and $1.0$ is the initial distance of the
vortex center from the beginning of the inlet. We test the vortex advection
with the methods and compare them with the results for a long domain. In
Figure~\ref{fig:vortex_pu}, we have plotted the pressure and velocity for the
different methods. In Table~\ref{table:vortex_l2} the $L_2$ errors of the
pressure and velocity are shown. It is evident from the plots that, the
do-nothing, modified do-nothing, and our new hybrid method match the results
of a long domain well. However, in case of mirroring technique a lot of back
pressure fluctuation is visible. The MOC shows a significant deviation from
the long domain and also does not preserve the velocity variation. In the
pressure plot, we can see that the mirror method shows a perfect match before
the vortex reaches the probe, thus it is suitable for outlets with very low
gradients. However, once the vortex reaches the probe, the results of the
mirror method are very poor. The $L_2$ errors clearly show that the do-nothing
methods and the hybrid schemes work well. In particular, the error in $p$ and
$v$ is high for the mirror and method of characteristic.
\begin{table}[h!]
\centering
\begin{tabular}{lrrr}
\toprule
        Methods & $e(p^{*})$ & $e(u^{*})$ & $e(v^{*})$ \\
\midrule
 Characteristic &      1.216 &      0.013 &      0.959 \\
     Do-Nothing &      0.554 &      0.002 &      0.174 \\
         Hybrid &      0.569 &      0.002 &      0.173 \\
         Mirror &      1.588 &      0.010 &      0.806 \\
 New Do-Nothing &      0.629 &      0.002 &      0.182 \\
\bottomrule
\end{tabular}

\caption{$L_2$ error in the $p^*, u^*$, and $v^*$ measured at the probe for
  the moving vortex problem.}
\label{table:vortex_l2}
\end{table}

As mentioned earlier, all our simulations are inviscid which suggests an
infinite Reynolds number. However, to investigate the effect of Reynolds
number on the outlets we performed the above simulation at $Re=100$ and
$10000$. In Table~\ref{table:vortex_re100_l2} and
\ref{table:vortex_re10000_l2}, we show the errors for $Re=100$ and $10000$
respectively. We can clearly see that the hybrid method and do-nothing
have low errors compared other methods.  We can also see that as the Reynolds
number reduces and the fluid becomes increasingly viscous that the errors in
the method of characteristics as well as the mirror method reduce.  This
clearly shows the importance of the new method.

\begin{table}[h!]
\centering
\begin{tabular}{lrrr}
\toprule
        Methods & $e(p^{*})$ & $e(u^{*})$ & $e(v^{*})$ \\
\midrule
 Characteristic &      0.556 &      0.001 &      0.263 \\
     Do-Nothing &      0.521 &      0.001 &      0.258 \\
         Hybrid &      0.519 &      0.001 &      0.255 \\
         Mirror &      0.475 &      0.006 &      0.725 \\
 New Do-Nothing &      0.556 &      0.001 &      0.263 \\
\bottomrule
\end{tabular}

\caption{$L_2$ error in the $p^*, u^*$, and $v^*$ measured at the probe for
  the moving vortex problem, with $Re=100$.}
\label{table:vortex_re100_l2}
\end{table}
\begin{table}[h!]
\centering
\begin{tabular}{lrrr}
\toprule
        Methods & $e(p^{*})$ & $e(u^{*})$ & $e(v^{*})$ \\
\midrule
 Characteristic &      1.213 &      0.013 &      0.961 \\
     Do-Nothing &      0.553 &      0.002 &      0.175 \\
         Hybrid &      0.566 &      0.002 &      0.173 \\
         Mirror &      1.541 &      0.010 &      0.803 \\
 New Do-Nothing &      0.589 &      0.002 &      0.181 \\
\bottomrule
\end{tabular}

\caption{$L_2$ error in the $p^*, u^*$, and $v^*$ measured at the probe for
  the moving vortex problem, with $Re=10000$.}
\label{table:vortex_re10000_l2}
\end{table}

\subsection{2D backward-facing step}
\label{sec:bfs}

We consider the 2D backward-facing step problem. Following the experimental
work of \citet{Armaly1983}, the step height is set as, $h=4.9mm$ with inlet
width $h_{i}=5.2mm$. We compare the velocity profile at different stations
with $x/h= 2.55, 3.57, 4.80, 7.14$ (where $x$ is the distance downstream from
the step). We compare our results with the experimental results in
\cite{Armaly1983}. The Reynolds number of the flow is chosen to be $389$ since
above this the flow is no longer two-dimensional. In the simulation, we set
$\rho=1.225kg/m^3$ and the viscosity is calculated using $Re=2 \bar{U} h / \nu
$, where $\bar{U}=2/3 U_{max}$ is the mean velocity. The inlet velocity is set
to $1 m/s$. The schematic of the simulation model is shown below

\begin{figure}[h!]
	\centering
  \includegraphics[width=\textwidth]{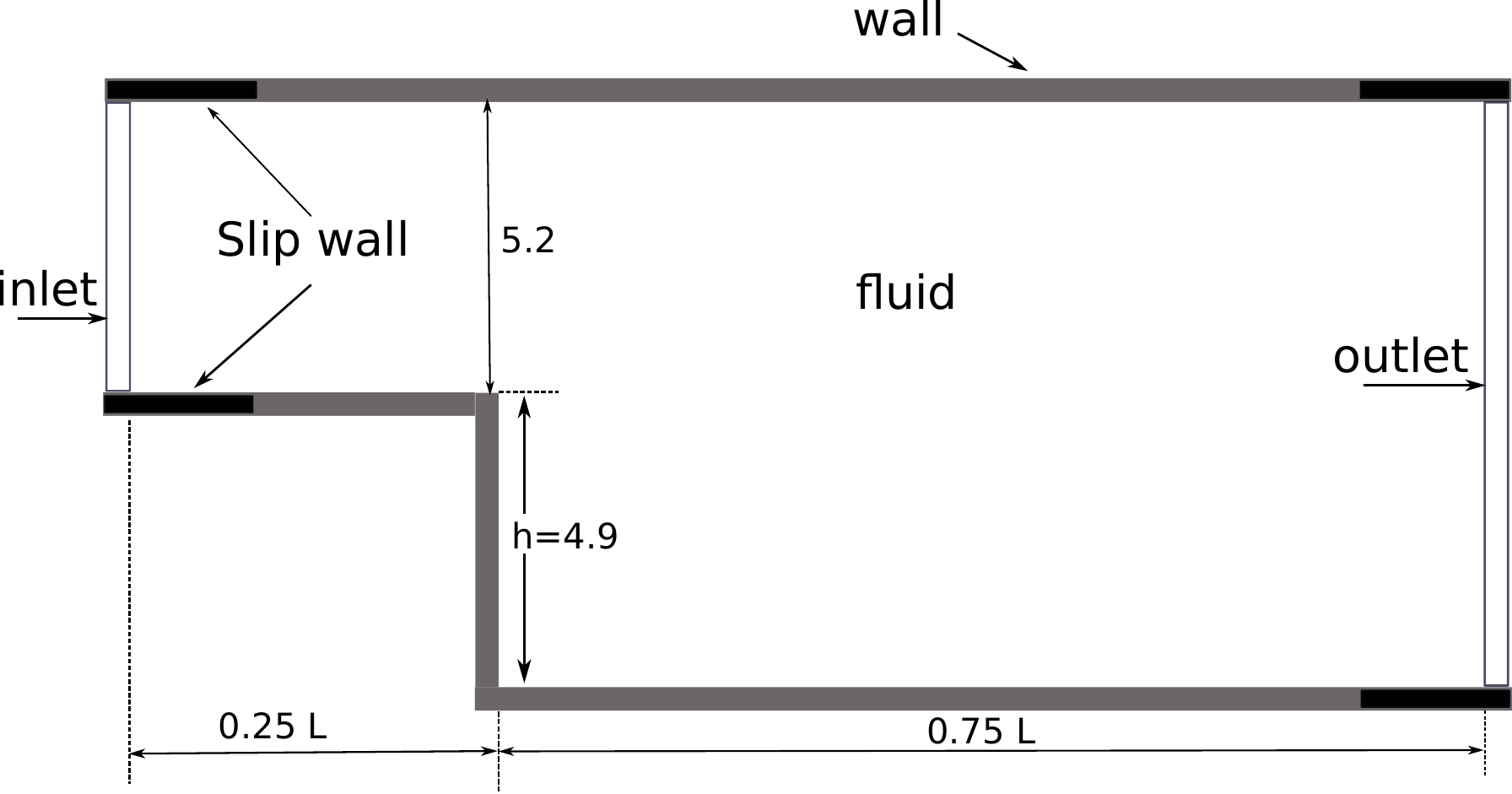}
	\caption{Sketch of domain used for backward-facing step simulations (all
    dimensions in mm).}
	\label{fig:bfs}
\end{figure}

\begin{figure}[ht!]
	\centering
  \includegraphics[width=0.8\textwidth]{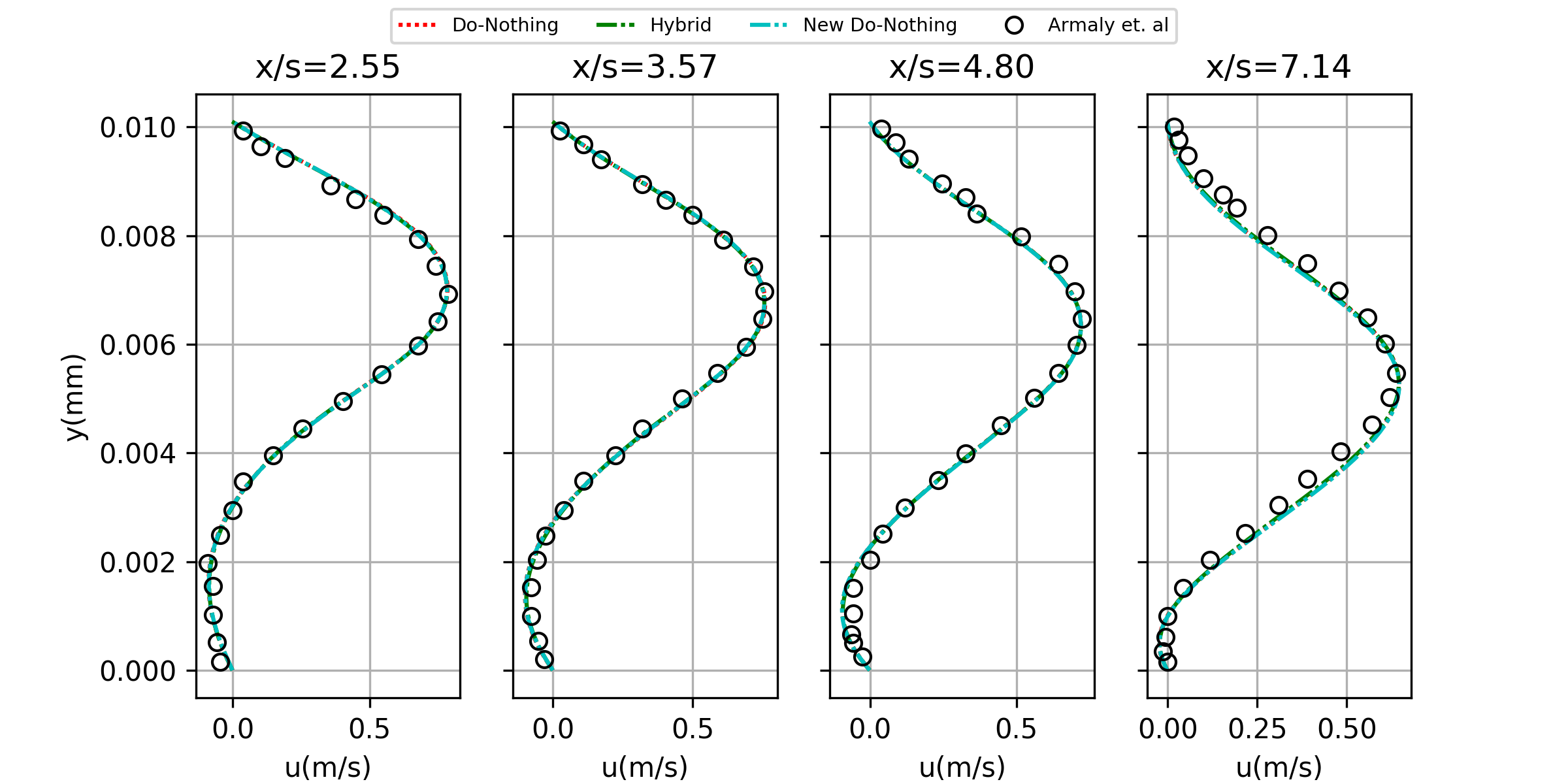}
	\caption{Velocity at $t = 1 sec$ for $Re=389$ at different locations.}
	\label{fig:bfs_u}
\end{figure}

At the walls, we satisfy the no-slip boundary condition. However, since the
inlet is set at a constant velocity, a no-slip wall introduces non-physical
pressure fluctuations. Thus a small part of the initial wall is set as a slip
wall. Similarly, near the outlet we allow slip at the wall in order to avoid
vortices at the start of the flow. In this test, we have shown results for our
proposed method and do-nothing only, since the characteristic method and
mirror methods failed to complete. In case of the mirror method, the vortices
reach the outlet and the simulation blows up. In case of characteristics, the
criteria for reference parameter is not known. In Figure~\ref{fig:bfs_u}, we
show the velocity profile for all the methods. It is evident from the plot
that all the methods (hybrid, do-nothing, and modified do-nothing) are able to
reproduce the results presented by \cite{Armaly1983}.

\begin{table}[h!]
\centering
\begin{tabular}{lr}
\toprule
         Method & $x_{rl}/h$ \\
\midrule
     Do-Nothing &      8.030 \\
         Hybrid &      8.009 \\
 New Do-Nothing &      7.901 \\
\bottomrule
\end{tabular}

\caption{The reattachment length for $Re=389$ for different outlet implementations.}
\label{table:bfs_rl}
\end{table}

The reattachment length for the primary vortex is determined and presented in
Table~\ref{table:bfs_rl}. We can clearly see that the reattachment length is very
close to the experimental value $7.94$ from \cite{Armaly1983}. This testcase clearly
shows that the proposed method shows very less difference from the
experimental values compared to other methods. This also highlights the ability
of new proposed testcase to distinguish between truly non-reflecting outlet boundaries.

\subsection{Flow past circular cylinder}
\label{sec:cylinder}

\begin{figure}[h!]
	\centering
  \includegraphics[width=\textwidth]{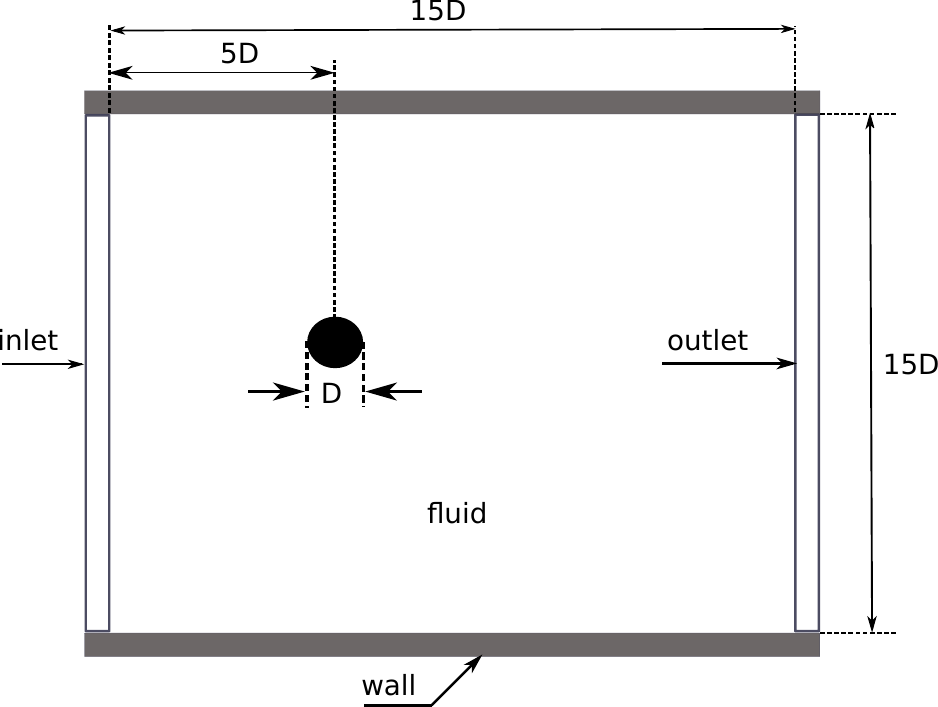}
	\caption{Sketch of domain used for flow past circular cylinder simulations.}
	\label{fig:fpc}
\end{figure}

\begin{figure}[ht!]
	\centering
  \includegraphics[width=0.8\textwidth]{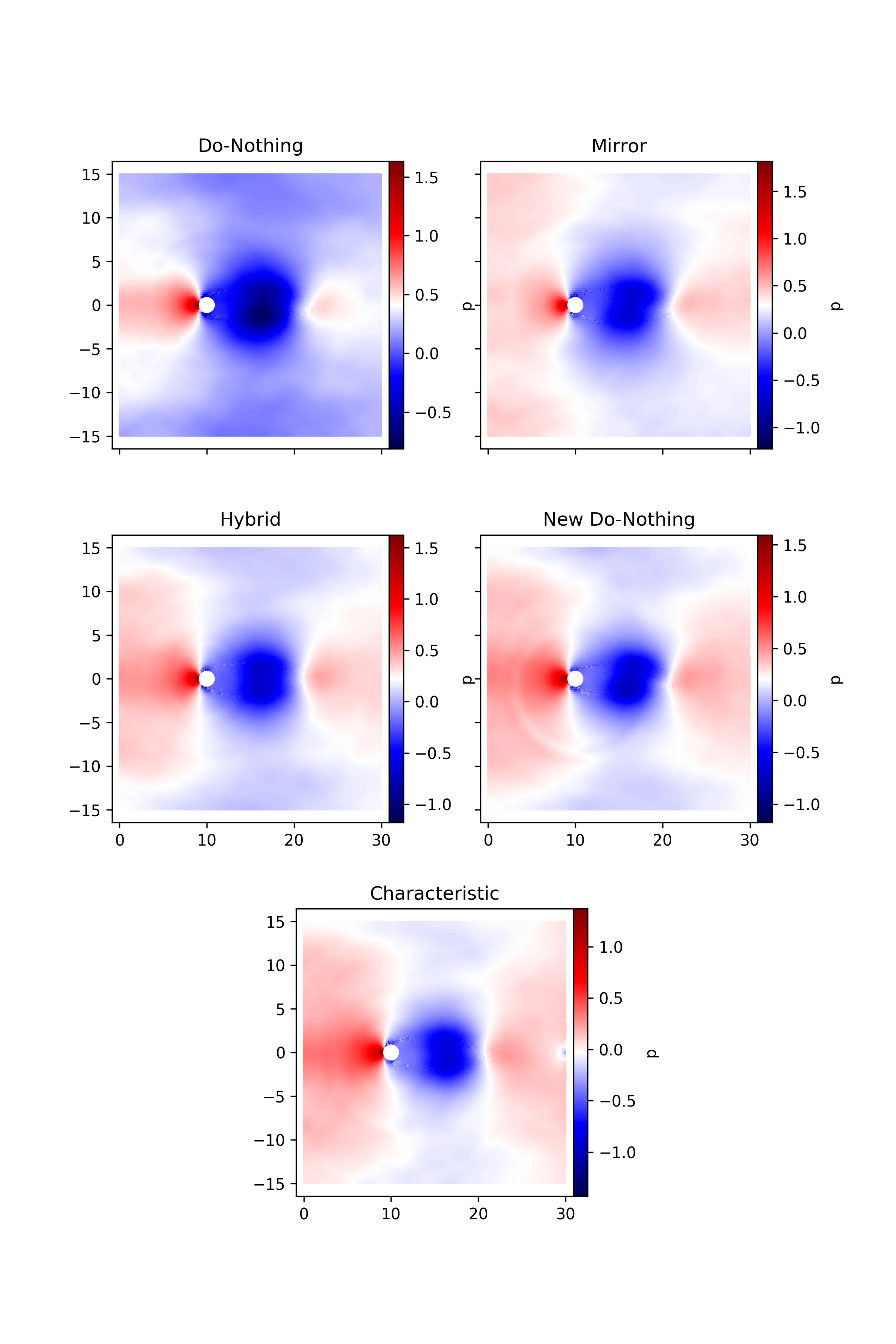}
	\caption{Normalized pressure at $t = 50 sec$ for $Re=200$.}
	\label{fig:pplots}
\end{figure}

\begin{figure}[ht!]
	\centering
  \includegraphics[width=0.8\textwidth]{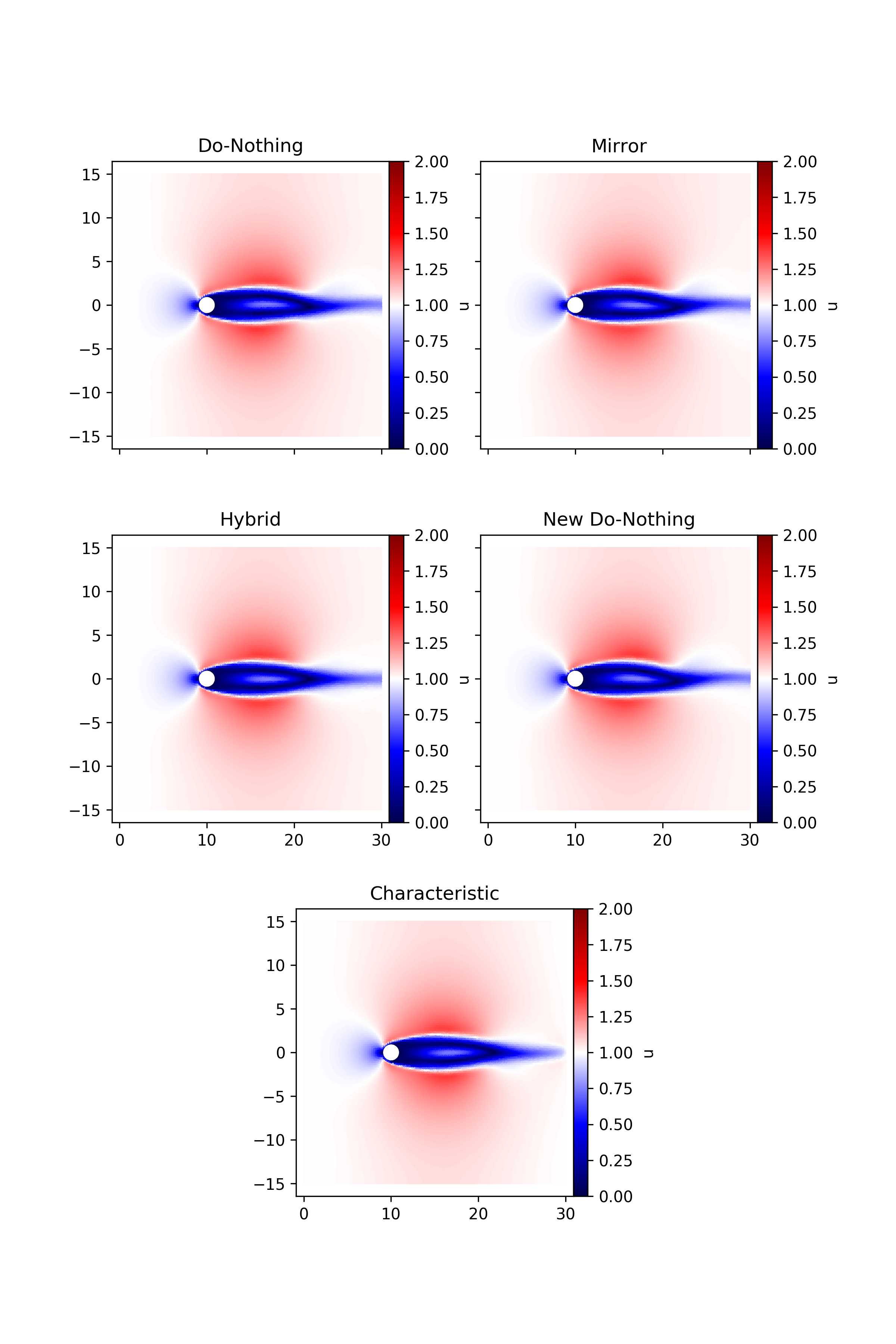}
	\caption{Normalized velocity at $t = 50 sec$ for $Re=200$.}
	\label{fig:vplots}
\end{figure}

\begin{figure}[ht!]
	\centering
  \includegraphics[width=0.8\textwidth]{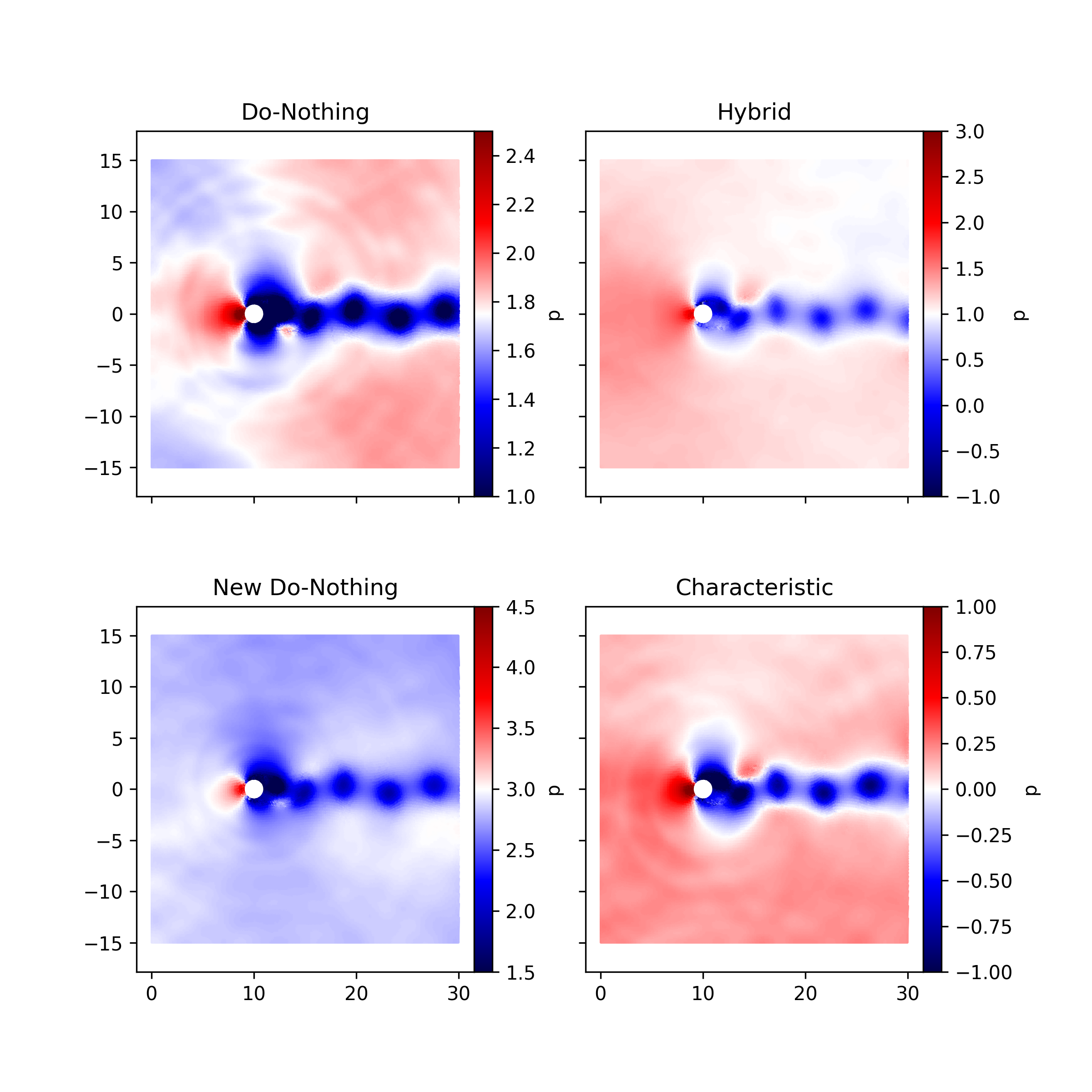}
	\caption{Normalized pressure at $t = 150 sec$ for $Re=200$.}
	\label{fig:pplots1}
\end{figure}

\begin{figure}[ht!]
	\centering
  \includegraphics[width=0.8\textwidth]{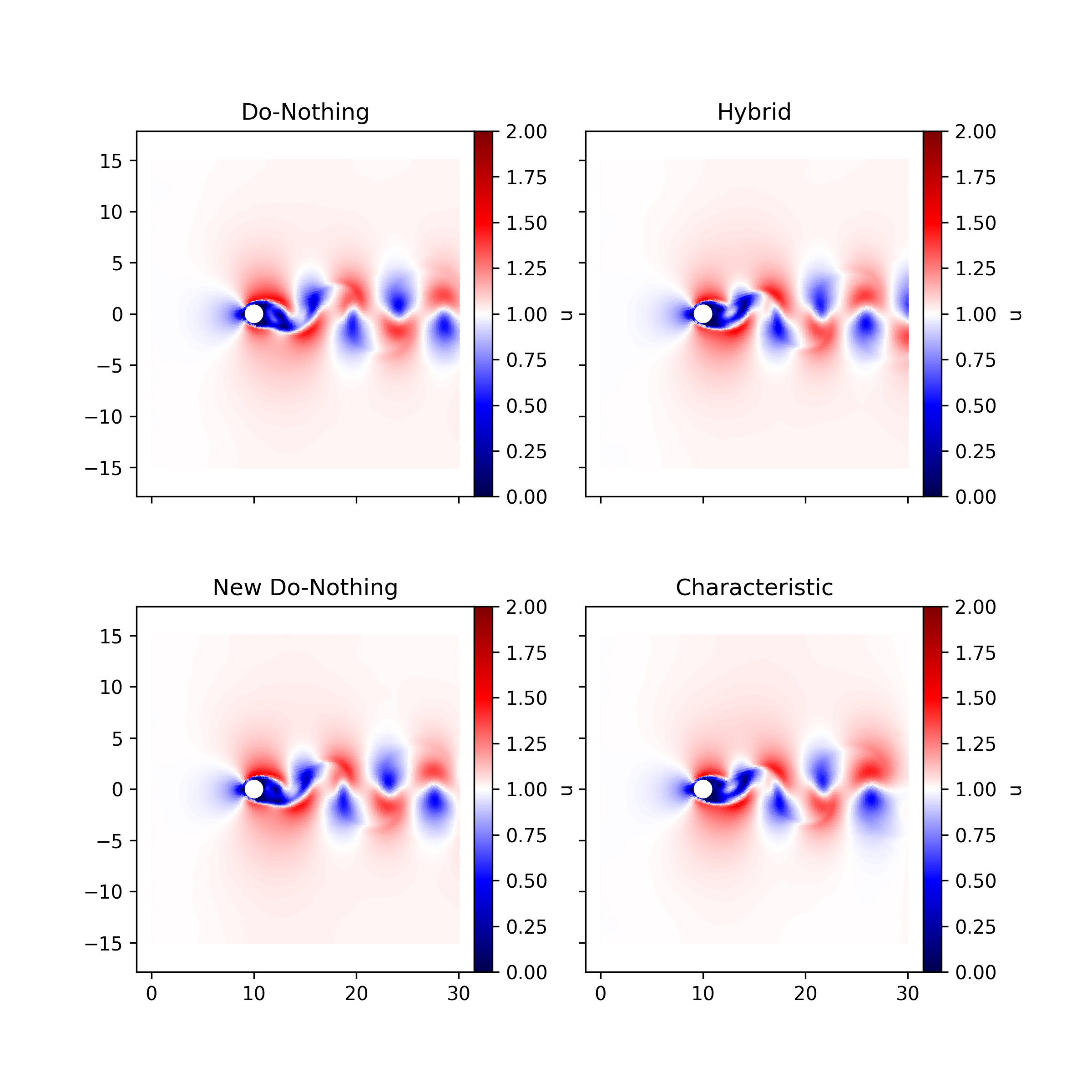}
	\caption{Normalized velocity at $t = 150 sec$ for $Re=200$.}
	\label{fig:vplots1}
\end{figure}
The flow past a circular cylinder is a well known benchmark to show the
capability of inlet/outlet boundaries. We investigate the problem for all the
methods described in this paper. We consider a smaller domain compared to
earlier research with fewer particles to show the effectiveness of the
proposed method~\cite{open_bc:tafuni:cmame:2018, MARRONE2013456}. A cylinder
of diameter $D (=2m)$ has been considered. The channel width is $15D$ to avoid
the effect of wall and the length is $15D$, which is aligned along the x-axis.
The cylinder is at $5D$ from the inlet interface as shown in Figure
\ref{fig:fpc}. Each inlet, outlet and wall has $6$ layers of particles which
are enough to get full kernel support. The inlet is given a constant
prescribed velocity, $u_\infty=1 m/s$. The walls function as a slip wall in
order to avoid effect of boundary layer from the walls. The fluid properties
such as kinematic viscosity of the flow is evaluated using $\nu = u_\infty
D/Re$, where $Re$ is the Reynolds number of the flow and density $\rho = 1000
kg/m^3$. We use a particle spacing $\Delta x=0.0667$ and $h/\Delta x=1.2$
which result in $201694$ fluid particles in the domain. This spacing results
in a cell Reynolds number of $Re_{cell}=u_\infty h/\nu$ of 8. This suggests
that this is a coarse simulation. In order to capture the curvature of the
cylinder, we place the particles in the solid such that the volume is
consistent. We first place particles on the circumference spaced $\Delta x$
from each other and then create points on a circle $\Delta x$ towards center
and perform the same procedure until we reach the center.

We simulate the model for $Re=200$ for all the methods. In the
Fig.~\ref{fig:pplots} and \ref{fig:vplots} we have plotted $p^*$ and $u^*$
respectively at $t=50s$ for all methods. Since the gradient near the outlet
boundary is close to zero, all the methods show similar variations. However
after vortex shedding starts, the gradient near the outlet is large. In
Fig.~\ref{fig:pplots1} and \ref{fig:vplots1} we show the pressure and velocity
distribution at $t=150s$ respectively, when the vortex shedding is well
established. In case of mirror method due to high gradient near the outlet,
spurious pressures arise and the particle positions diverge. It is evident
from the pressure plots in Fig.~\ref{fig:pplots1} that the MOC reflects the
pressure back into the domain when vortex shedding starts. In case of
do-nothing and modified do-nothing a significant increase in pressure of the
domain is visible. Pressure for both hybrid and characteristic method looks to
be distributed around zero which is essential for low numerical errors in
pressure calculations. In Fig.~\ref{fig:vplots1} of velocity distributions,
all the methods show a similar pattern and it is hard to comment on the
relative
merits of the methods.\\
\begin{figure}[h!]
	\centering
  \includegraphics[width=\textwidth]{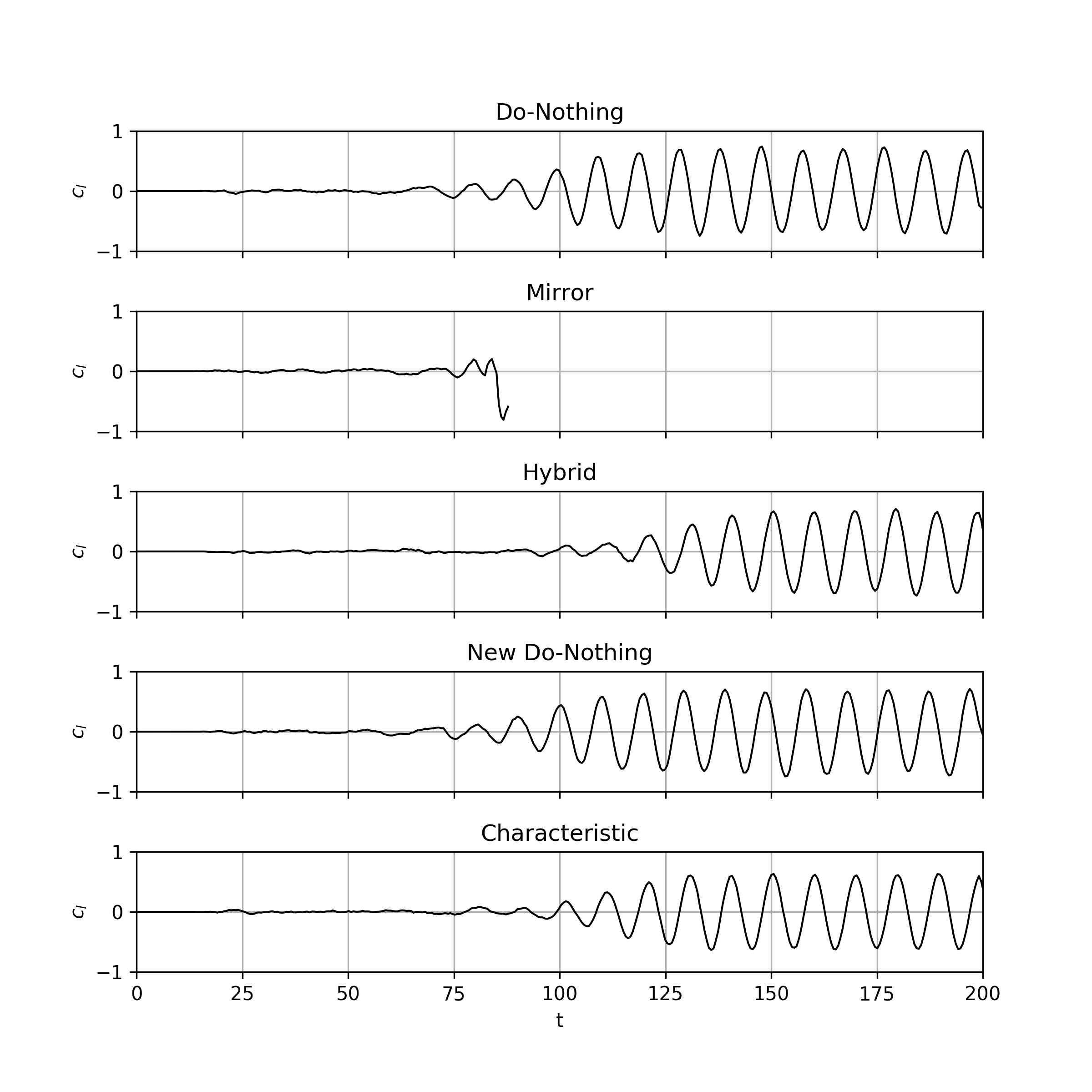}
  \caption{Plot for $c_d$ for all methods at Re=200}
	\label{fig:cdplots}
\end{figure}
In order to check the accuracy of the methods, we calculate the drag
$(F_d=F_x)$ and lift ($F_l=F_y$) forces on the cylinder for all the cases and
evaluate the coefficient of drag, $c_d=F_d / (0.5 \rho u_{\infty}^{2})$, and
lift $c_l=F_l / (0.5 \rho u_{\infty}^{2})$. A five point average is taken to
filter the noise. The force on the solid cylinder is determined by solving the
momentum equation given by
\begin{align}
	\label{eq:mom-newt_solid}
  \frac{F_{solid}}{m_{solid}} &= -\frac{1}{\rho} \nabla p +
  \nu \nabla^2 \ten{u}.
\end{align}
The above equation in the SPH form is given by
\begin{equation}
	\label{eq:tvf-momentum_solid}
	\begin{split}
    F_{solid} = \sum_j \left( V_i^2 +
		V_j^2 \right) & \left[ - \tilde{p}_{ij} \nabla W_{ij} +
      \tilde{\eta}_{ij} \frac{\ten{u}_{ij}}{(r_{ij}^2 + \eta
    h_{ij}^2)} \nabla W_{ij}\cdot \ten{r}_{ij} \right],
	\end{split}
\end{equation}
where all symbols have same meaning as given in section \ref{sec:sph} except
$u_{ij} = u_{g_i} - u_{j}$, where $u_{g_{i}}$ is the solid wall velocity
\cite{zhang_hu_adams17}. We also evaluate the Strouhal number $St=fD/u_\infty$ where
$f$ is the frequency of shedding. In Fig.~\ref{fig:cdplots}, we compare $c_l$
for all the methods over time. It can be easily seen that mirror method blows
up after a large back pressure. In case of do-nothing and modified do-nothing,
shedding starts earlier compared to hybrid and characteristic method. In
Table~\ref{table:clcd}, we compare $c_d$ and $c_l$ and $St$ for all the
methods and with results published by \cite{guerrero2009numerical,
  MARRONE2013456, open_bc:tafuni:cmame:2018}. We can see that in spite of
having non-physical pressure variations in characteristic methods the value of
$c_d$ and $c_l$ shows a close match. In case of both do-nothing and modified
do-nothing the values are close since the pressure increase of the domain is
insignificant in case of incompressible flows. In our proposed hybrid method,
the pressure and velocity plots looks similar to
\cite{open_bc:tafuni:cmame:2018, MARRONE2013456}, the $c_d$ and $c_l$ are in
acceptable range presented in literature.
\begin{table}[h!]
\centering
\begin{tabular}{lrrr}
\toprule
                            Method &               $c_d$ &         $c_l$ &           St \\
\midrule
                    Characteristic &    $1.494 \pm 0.05$ &   $\pm 0.634$ &  $\pm 0.200$ \\
                        Do-Nothing &    $1.532 \pm 0.05$ &   $\pm 0.744$ &  $\pm 0.210$ \\
                            Hybrid &    $1.524 \pm 0.05$ &   $\pm 0.722$ &  $\pm 0.210$ \\
                    New Do-Nothing &    $1.540 \pm 0.05$ &   $\pm 0.729$ &  $\pm 0.210$ \\
            \citet{MARRONE2013456} &     $1.38 \pm 0.05$ &  $ \pm 0.680$ &     $ 0.200$ \\
     \citet{guerrero2009numerical} &  $1.409 \pm 0.048 $ &  $ \pm 0.725$ &            - \\
 \citet{open_bc:tafuni:cmame:2018} &              $1.46$ &  $ \pm 0.693$ &     $ 0.206$ \\
\bottomrule
\end{tabular}

\caption{Comparison of $c_l$, $c_d$ and $St$ values for different method with
literature for $Re=200$.}
\label{table:clcd}
\end{table}
Furthermore, the proposed hybrid and modified do-nothing methods have been tested
for Reynolds number 20.
\begin{figure}[h!]
	\centering
  \includegraphics[width=\textwidth]{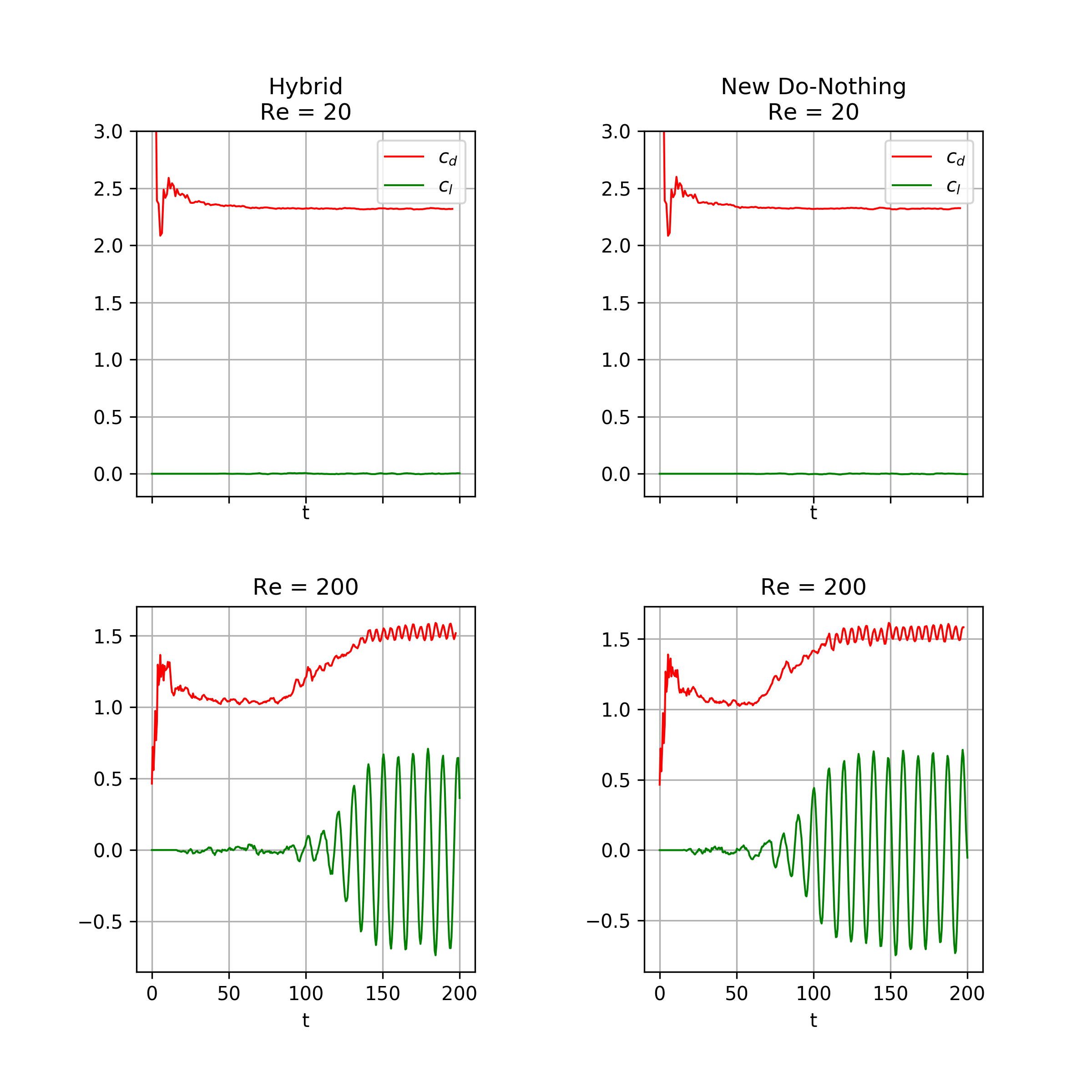}
  \caption{Plot for $c_d$ and $c_l$ for hybrid and the modified do-nothing at
    Re=20 and 200.}
	\label{fig:cdclplots}
\end{figure}
\begin{table}[h!]
\centering
\begin{tabular}{lrrr}
\toprule
Particle Spacing &   Hybrid & New do-nothing \\
\midrule
            0.05 &  $2.317$ &        $2.317$ \\
            0.07 &  $2.320$ &        $2.321$ \\
            0.10 &  $2.317$ &        $2.319$ \\
\bottomrule
\end{tabular}

\caption{Convergence of $c_d$ with the decrease in particle
  spacing at Re=20 for hybrid method}
\label{table:hybridconv}
\end{table}
In the Fig.~\ref{fig:cdclplots}, we show the $c_d$ and $c_l$ for hybrid and
modified do-nothing. It is evident from the figure that both modified
do-nothing and hybrid produces similar results. However hybrid is better than
the modified do-nothing as shown in other test cases. In order to check the
convergence of the results, we perform a convergence study for both the
proposed methods and tabulated the results in Table~\ref{table:hybridconv} at
$Re=20$. We observe that $c_d$ decreases with decrease in particle spacing and
converges to around 2.317. When the spacing is $0.05$ the cell Reynolds number
is 0.6 suggesting a sufficiently resolved simulation. It must be noted that
the results presented are in a smaller domain and with much fewer particles
than those used in earlier research which show that the proposed methods
replace need of a large domain for wind-tunnel type of simulations. The
results above also show that the flow past a circular cylinder does not reveal
important differences between the different boundary conditions and the
importance of our proposed test problems.


\section{Conclusions}
\label{sec:conclusions}

In this paper we review the established techniques for implementing outlet
boundaries in the context of weakly-compressible SPH schemes. We classify them
into three broad categories. In order to systematically examine these, we
construct four simple test problems. These tests clearly show the deficiencies
of the existing approaches.
\begin{itemize}
\item{Do-nothing method is only suitable for problems where high intensity
    acoustic pressure waves are absent.}
\item {The mirror method works best for flows where the gradients are very low
    near the outlet.}
\item {The MOC show excellent results where reference properties are known a
    priori but are not very effective when there are gradients in the flow at
    the exit.}
\end{itemize}
Based on this, we propose a new generalized scheme which combines the
do-nothing and characteristic based outlet into a new hybrid technique. The
proposed technique works well with both high intensity acoustic waves and high
gradient flow near the outlet. Unlike the MOC, it calculates reference flow
variables by time averaging. We also propose a much simpler and slightly
modified do-nothing boundary condition that produces good results. We then
demonstrate these with simulations of the flow past a circular cylinder at two
different Reynolds numbers and also for the flow past a backward-facing step.
We are able to obtain very good results with much fewer particles than
reported earlier. Finally, our implementation is open source and our
manuscript is fully reproducible.

\section*{Acknowledgements}

The authors are grateful to Prof.~Krishnendu Haldar of the Department of
Aerospace Engineering, IIT Bombay for providing us with his
workstation to expedite our simulations.

\section*{References}
\bibliographystyle{model6-num-names}
\bibliography{references}

\end{document}